\documentclass[prc,aps,amsfonts,amsmath,superscriptaddress,twocolumn,nopacs,floatfix]{revtex4}
\usepackage{footnote}
\usepackage[colorlinks=true]{hyperref}
\usepackage{graphicx} 
\usepackage{float} 
\usepackage{epsfig} 
\usepackage{subfig}
\usepackage{rotating}
\usepackage{xcolor}
\usepackage{braket}

\begin{document}

\title{Alpha-particle condensation: a nuclear quantum phase transition}
\author{J.-P. Ebran}\email{jean-paul.ebran@cea.fr}
\affiliation{CEA,DAM,DIF, F-91297 Arpajon, France}
\author{M. Girod}\email{michel-g.girod@cea.fr}
\affiliation{CEA,DAM,DIF, F-91297 Arpajon, France}
\author{E. Khan}\email{khan@ipno.in2p3.fr}
\affiliation{Institut de Physique Nucl\'eaire, Universit\'e Paris-Sud, IN2P3-CNRS, 
Universit\'e Paris-Saclay, F-91406 Orsay Cedex, France}
\author{R.D. Lasseri}\email{raphael.lasseri@cea.fr}
\affiliation{ESNT, CEA, IRFU, D\'epartement de Physique Nucl\'eaire,
Universit\'e Paris-Saclay, F-91191 Gif-sur-Yvette}
\author{P. Schuck}\email{schuck@ipno.in2p3.fr}
\affiliation{Institut de Physique Nucl\'eaire, Universit\'e Paris-Sud, IN2P3-CNRS, 
Universit\'e Paris-Saclay, F-91406 Orsay Cedex, France}

\begin{abstract}
When the density of a nuclear system is decreased, homogeneous states undergo the so-called Mott transition towards clusterised states, e.g. alpha clustering, both in nuclei and in nuclear matter.
Here we investigate such a quantum phase transition (QPT) by using microscopic energy density functional (EDF) calculations both with the relativistic and the Gogny approaches on the diluted $^{16}$O nucleus. The evolution of the corresponding single-particle spectrum under dilution is studied, and a Mott-like transition is predicted at about 1/3 of the saturation density. Complementary approaches are used in order to understand 
this QPT. A study of spatial localisation properties as a function of the density allows to derive a value of the Mott density in agreement with the one obtained by fully microscopic calculations in $^{16}$O and in nuclear matter. Moreover a study of the spontaneous symmetry breaking of the rotational group in $^{16}$O, down to the discrete tetrahedral one, provides further insight on the features displayed by the single-particle spectrum obtained within the EDF approach.The content of the tetrahedrally deformed A-nucleon product state in terms of spherical particle-hole configurations is investigated. Finally a study of quartet condensation and the corresponding macroscopic QPT is undertaken in infinite matter.
\end{abstract}


\date{\today}

\maketitle


\section{Introduction}
\label{Sec:Intro}

Fermi systems are the host of various phenomena
yet to be fully explored. 
One of the most recent exciting features which has been revealed is the exotic arrangements stabilized by the existence of internal degrees of freedom in N-component Fermi systems with $\text{N}>2$~\cite{wil09,nis12,wu03,cap08}. Molecular configurations made of bound states of N fermions enrich the celebrated crossover~\cite{bou04} between a Baarden, Cooper and Schrieffer (BCS)
superfluid phase to the  Bose-Einstein condensation (BEC) of bosonic 
bound states of two fermions that characterizes 2-component Fermi gases with an attractive s-wave interaction.  
Nucleons being assigned to spin and isospin
SU(2) doublets, atomic nuclei fall in the category of 4-component self-bound Fermi systems.
Attractive s-wave interactions
in the singlet-even
($S=0,T=1$) and the triplet-even ($S=1,T=0$) channels --- $S$ and $T$ stand respectively for the total spin and isospin momenta of the two-nucleon system,  with almost similar strength gives rise to 
various types of superfluid behavior. In the weak coupling regime, the dominant superfluid instability
manifests itself through the establishment of a 
BCS quasi-long range order and involves proton-proton, neutron-neutron or proton-neutron (depending on the matching of neutron and proton Fermi levels)
Cooper pairs~\cite{gez11,fra14}. Moving towards the strong coupling regime,  calculations in infinite symmetric nuclear matter~\cite{rop82}
suggested that the dominant superfluid 
order is not a BEC phase of bosonic dimers (deuterons), but rather a condensation phase of quartets --- 
4-fermion molecular objects with zero total spin and isospin.
Infinite nuclear matter hence undergoes for decreasing densities a phase transition to alpha-particle condensation \cite{rop82,rop98,hor06}. That is nuclear matter lowers its energy by taking advantage of the nuclear cohesion, i.e by forming localized clusters that recover saturation density ($\rho_0\sim 0.16\ \text{fm}^{-3}$ ), rather than remaining in a dilute homogeneous phase. Since this happens at zero temperature, it can be qualified as a Quantum Phase Transition (QPT) where the density is the control parameter. 

How such features translate in finite nuclei triggered several research works, see e.g.~\cite{toh01,yam04,yam12,gir13,ebr14}. Unlike homogeneous systems, finite nuclei display large fluctuations of their mass density around the equilibrium value $\rho_0$ either in the ground state of heavy nuclei, where the density near the surface gets shallower, or in excited states. In these local low-density regions, a QPT from a dilute homogeneous phase to a clusterized one is expected to occur, causing a preformation of alphas at the surface of heavy nuclei~\cite{typ14,ebr18} and endowing the spectroscopy of relatively light nuclei with clusterized excitation modes~\cite{fre07,fre18,mar18,mar19}.
For instance the famous Hoyle state, important for $^{12}$C production in the Universe, could be interpreted as a three-alpha gas state where the alphas occupy with their center of mass (c.o.m.) motion to 70-80 percent the lowest 0S wave function while all other states have an occupation probability more than by a factor ten down, see, e.g., ~\cite{sch13}. In that sense, the Hoyle state could be qualified as a finite-size alpha-particle condensate. However EDF and geometrical approaches end up with alpha-clusters in a much more robust configuration \cite{ebr14b,zha16,bij14}, see also~\cite{bis19} for a recent experimental investigation of this issue and~\cite{ohk19} for a discussion of alpha-cluster structures as a manifestation of supersolidity. It should be noted that the action of the Pauli principle is quite similar in both cases (gas or molecular states) so converging results shall be reached from both approaches. Beyond these interpretations, the size of the Hoyle state is extended to 3-4 the volume of the $^{12}$C ground state~\cite{fun03,che07}, showing that Hoyle and ground states live in two completely different phases, one dilute, the other dense.

In this work, we want to further substantiate  the QPT scenario of alpha-clustering for the case of $^{16}$O through complementary perspectives. Section~\ref{Sec:Loc} first describes how relevant dimensionless quantities such as the localisation and the Brueckner parameters help characterizing the transition from a homogeneous nuclear system towards a localized one. A microscopic analysis based on the energy density functional (EDF) approach is given in section~\ref{Sec:EDF}, where both covariant and Gogny functionals are used. Constrained Hartree-Fock-Bogoliubov (HFB) calculations for $^{16}$O are performed, the constraint being on the radius of $^{16}$O while the system is imposed to stay globally spherical. That is $^{16}$O can break up into clusters while the system still stays spherical on average. The single-particle properties are then addressed in the light of group theory and molecular orbitals.  
Finally section~\ref{Sec:Schuck} provides an analysis of the 
occurrence of alpha-condensation in nuclear matter through the explicit treatment of four-nucleon correlations.

\section{Localisation and quantum phase transition in nuclear systems}
\label{Sec:Loc}
In contrast with ultra-cold atom physics, it is not possible to directly tweak the effective
strength of the interaction between nucleons, for nuclei being self-bound
systems. On the other hand, as we shall see below, nucleon density is one
of the control parameters driving nuclei from the weak to strong coupling regimes. 
The nuclear saturation density $\rho_\text{sat}\sim\ 0.16\text{fm}^{-3}$ is an emergent property~\cite{hag16} 
setting the characteristic energetic and spatial scales in nuclear systems.  
Finite nuclei display large fluctuations around $\rho_\text{sat}$, such that nucleons evolving 
in the low-density parts of finite nuclei enter a strong coupling regime and can hence self-arrange
into alpha-clusters.

The analysis of fundamental dimensionless parameters sheds light on how 
nuclear density drives the coupling regime of its constituents.  
Indeed, the essential features of nuclear systems are grasped by
dimensionless parameters defined in terms
of the characteristic energy and length scales of the problem.
For instance, the quantum mechanical nature of nuclear systems 
over a wide range of densities can be justified by expressing 
a first dimensionless parameter --- the localisation parameter
$\alpha_{loc} \simeq\frac{\lambda}{\bar{r}}$~\cite{ebr12,ebr13} ( with $\lambda$ the typical spatial extension of nucleons and $\bar{r}$ 
their mean interparticle-distance)
in terms of the nuclear density.
A value smaller or larger than 1 allows to discuss about localised or delocalized states in quantal systems, whereas 
the $\alpha_{loc}\ll 1$ limit corresponds to a classical system.
In the zero-temperature case and owing to the fact that nucleons are bound in a nucleus, $\lambda$ 
is well approximated by the confining length of an harmonic oscillator (HO) potential 
which parameters are chosen to match the radius $R$ of a given nucleus.
We then have in $\hbar=c=1$ units
\begin{equation}
\lambda \simeq \sqrt{\frac{1}{m\omega}}=\frac{\sqrt{R}}{(2mV_0)^\frac{1}{4}},
\end{equation}
with $\omega$ the typical energy of the HO, $m$ the nucleon mass
and $V_0$ the depth of the confining potential.
In order to study the behavior of the localisation parameter with the density of the system, and not only at saturation density,
let us  express the mean-interparticle distance as
$\bar{r} = \left(\frac{3}{4\pi\rho}\right)^\frac{1}{3}$.
One gets in terms of the average density $\rho$
\begin{equation}
\alpha_{loc} \sim \frac{(A\rho)^\frac{1}{6}}{(mV_0)^\frac{1}{4}},
\label{eq:al}
\end{equation}
where A is the number of consitutents of the system.
Larger densities or a shallower confining potential hence drives the system
towards the quantal regime. At saturation density, one typically obtains 
values of $\alpha_{loc}$ between 0.8 and 1.5 while at one tenth
of the saturation density, a factor 0.7 arises showing that  nuclear systems remain quantal
over a large range of densities.

It should be noted that behaving quantum-mechanically, the effective strength of nucleons
interaction, i.e. the extent to which interactions impact the properties of nuclei
and make the latter deviates from the ideal free case, is sometimes measured by another 
dimensionless ratio between the mean potential $\braket{V}$ and the mean quantal kinetic $\braket{K}$ energies,
or equivalently the ratio between the mean interparticle distance $\bar{r}$ and 
the generalized Bohr radius $a_B$. This is the so-called
Brueckner (also known as the Wigner) parameter (see e.g.~\cite{bon08})
$r_s=\frac{\braket{V}}{\braket{K}}\sim\frac{\bar{r}}{a_B}$. 
In the nuclear case, 
we have
\begin{equation}
r_s=\frac{\omega}{E_F},
\label{eq:rs}
\end{equation}
where the energy of the harmonic oscillator reads 
$\omega=\frac{1}{R}\sqrt{\frac{2U}{m}}$ and 
$E_F=\frac{1}{2m}\left(\frac{3\pi^2\rho}{2}\right)^\frac{2}{3}$
stands for the Fermi energy. One gets from Eqs. (\ref{eq:al}) and (\ref{eq:rs}):
\begin{equation}
r_s\sim\alpha_{loc}^{-2}
\end{equation}
showing that the localisation parameter captures the relevant effects. Three quantities 
govern these dimensionless parameters: i) the number of nucleons $A$, ii) the depth of the confining potential
$V_0$ and iii) the average density of the system $\rho$.
Nuclei are in the strong coupling regime when the Brueckner
parameter $r_s$ is large enough (hence the localisation parameter small enough). In other words, i) fewer nucleons, ii) a deeper confining potential and iii) lower
densities favor states where quartetting correlations dominate, i.e where clustering 
is likely to occur. The first two effects have been studied in~\cite{ebr12,ebr13,ebr14,ebr14b} and we focus here on the role played by the nucleon density.
Because of Pauli blocking effects, 
alpha-particles in the nuclear medium start dissolving as they overlap with each others. The critical (Mott) density
at which this happens can be deduced from the dimensionless parameter $\alpha_{loc}^\alpha$ where the ratio is now between 
the size of an alpha-particle $R_\alpha \sim r_0A_\alpha^\frac{1}{3}$
and the mean internucleon distance $\bar{r}$: it is solution
of $\alpha_{loc}^\alpha=1$, that is
\begin{equation}
\frac{\rho^\alpha_\text{Mott}}{\rho_0}=\frac{3}{4\pi r_0^3A_\alpha\rho_0}\sim 0.2.
\end{equation} 
Below $\sim \rho_0/5$, quartetting correlations are no more suppressed by Pauli blocking effects
and a full alpha-clustering should correspond to the favoured arrangement of alpha-conjugate 
nuclei. We now wish to further substantiate the broad picture given through the analysis of dimensionless parameters by 
investigating the QPT in the framework of the EDF approach.

\section{EDF approach to Quantum Phase Transition in finite nuclei}
\label{Sec:EDF}


\subsection{Clustering and deformation}

\begin{figure}[!hbt]
      \includegraphics[width=0.5\textwidth]{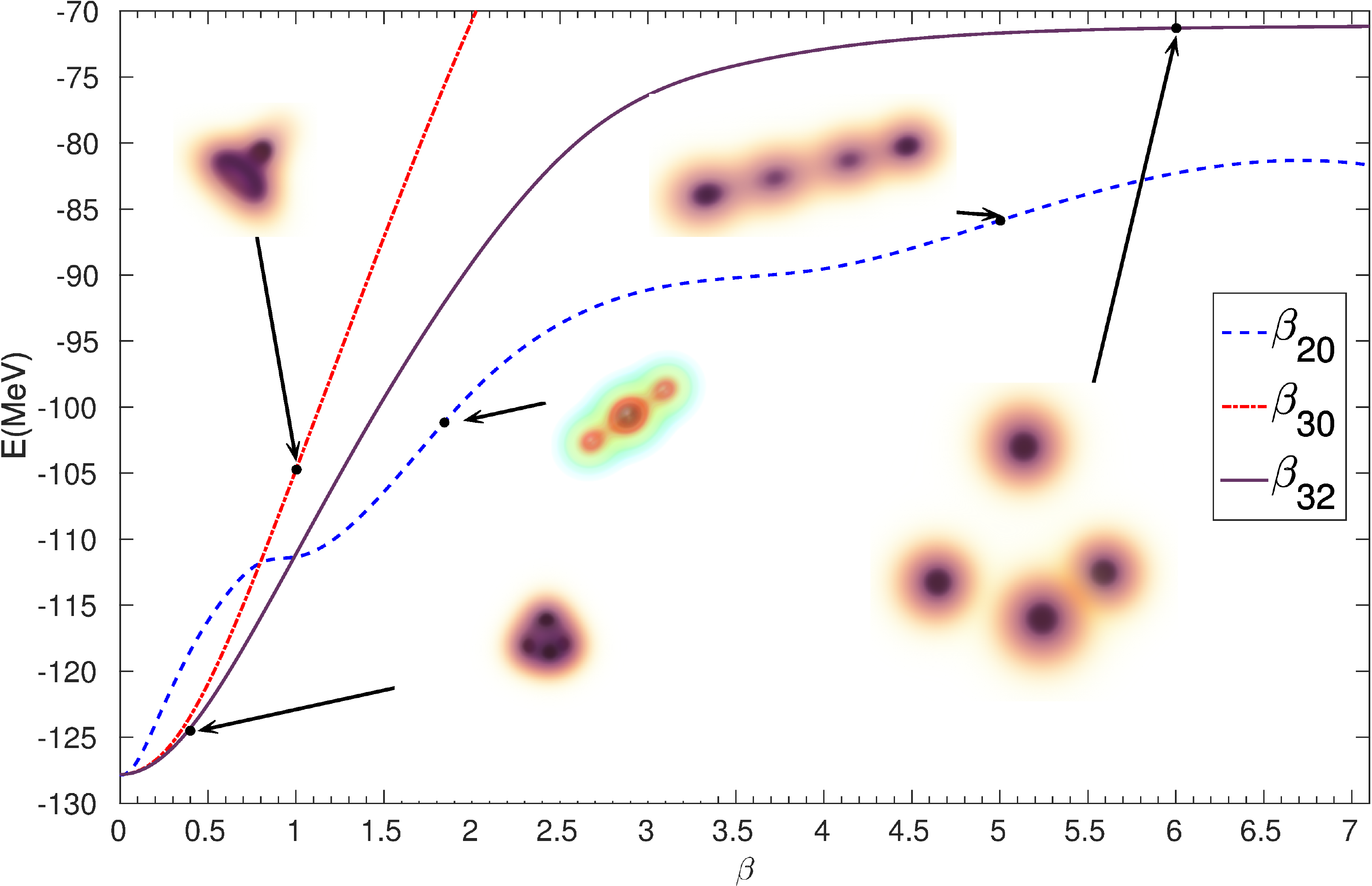}   
    \caption{(Color online) Binding energy of $^{16}$O as a
    function of a deformation parameter (axial quadrupole, axial octupole and tetrahedral ones), calculated within the CEDF at the SR level with the DD-ME2 parametrization~\cite{lal05}.The inserts display the 3D nucleonic density in the intrinsic frame of the nucleus for various values of the deformation parameter.}
    \label{fig:O16Evsbeta}
  \end{figure}

In the EDF spirit, rather than introducing explicit 4-body correlations,
it is natural  to look for an order parameter associated to alpha-clustering, 
i.e. a collective field whose fluctuations cause nucleons to gather into 
alpha-subunits. The multipolar mass moments $Q_{\lambda\mu}$ play such a role. 
Nuclear deformation indeed provides a necessary (yet not sufficient) condition for localized 
substructures to emerge in nuclei. The extreme scenario of 
all nucleons aggregating into alpha degrees of freedom corresponds to a spontaneous
breaking of the rotational symmetry $O(3)$ of the nuclear Hamiltonian
down to a discrete point group that dictates the geometrical configuration of the 
alpha-particles. Fig.~\ref{fig:O16Evsbeta} illustrates the relation between nuclear deformation and clustering by plotting  
$^{16}$O total binding energy computed at the Single-Reference (SR) level of the covariant energy density functional (CEDF) approach (also referred to as mean field level)
against constrained deformation parameters, namely the axial quadrupole 
$(\lambda,\mu) = (2,0)$, axial octupole $(\lambda,\mu) = (3,0)$ and  
triaxial octupole $(\lambda,\mu) = (3,2)$ modes. $^{16}$O intrinsic densities are also displayed for values of 
interest of the deformation parameters.
The global minimal energy is found at the spherical point, owing to the p-shell closure in $^{16}$O. 
Small values of the deformation parameters ($< 1$) correspond to deformed shapes where nucleons are roughly 
homogeneously distributed : prolate cigar-like shape
along $\beta_{20}$, pear-like configuration along $\beta_{30}$, tetrahedral distribution along $\beta_{32}$. 
The energy is rather stiff in the $\beta_{20}$ direction, contrary to the octupole directions, 
especially the triaxial one : for $\beta_{32} \sim 0.3$ where the tetrahedral shape is already well developed, 
the energy loss with respect to the spherical configuration is only 3 MeV. Projection on both angular momentum and parity 
as well as mixing within the generator coordinate method may therefore induce tetrahedral correlations in the ground state of $^{16}$O 
and yield several rotation-vibration excited states that can be classified according to the irreducible representations 
(irreps) of the discrete tetrahedral group $Td$, along the lines of~\cite{bij14}. For large deformation parameters, Fig.~\ref{fig:O16Evsbeta} involves binary cluster structures, e.g. $\alpha + ^{12}$C at $\beta_{30} \sim 1$, followed by ternary cluster structures ($^8Be+2\alpha$
at $\beta_{20}\sim 1.7$).
For extreme values of the deformation parameters where the nuclear radius is large, hence the average density low enough for Pauli blocking effects to be suppressed, $^{16}$O displays a fully clusterized structure with four alphas in linear ($\beta_{20}\sim 5$) or tetrahedral
($\beta_{32}\sim 6$) configurations.

For the sake of completeness, let us recall the link between deformation and cluster formation before 
investigating how the mean density drives alpha-clustering. 
The reason why deformation is intimately related to the occurrence of clusters can already be 
understood assuming a nuclear confining potential close to a HO one.
In the isotropic case (spherical configuration), the degeneracies of the energy levels of an N-dimensional HO are in one-to-one correspondence with 
the irreducible totally symmetric representations of $SU(N)$. Similarly, the quantum states of an N-dimensional anisotropic oscillator (deformed configuration) with commensurate frequencies (i.e. rationally related frequencies $\omega_i$ such that $k_i \omega_i = \omega$ and $k_i$ integral and relatively prime), specified by quantum numbers ${n_i}$ and possessing energies
\begin{equation}
E_{\left\lbrace{n_i}\right\rbrace} = \hbar\omega\sum_i\left(n_i+\frac{1}{2}\right)\frac{1}{k_i},
\end{equation}  
enjoy degeneracy spaces that also correspond to the representations of $SU(N)$, with the important difference that 
unlike the isotropic oscillator, a given representation occurs not singly but with a multiplicity $\prod_i k_i$~\cite{kin73,ros89}. Indeed,
the anisotropic HO with commensurate frequencies is unitarily equivalent to the direct sum of $\prod_i k_i$ isotropic
HOs with frequencies $\omega$, such that the phase space of the former can be mapped into the phase space of the isotropic HO at the 
cost of introducing a foliated (multi-sheeted) structure in the phase space~\cite{bha94}. For instance, a 3-dimensional anisotropic HO with axial symmetry and frequencies in a ratio $\omega_x=\omega_y\equiv\omega_\perp:\omega_z = 2:1$
(superdeformed prolate configuration), i.e. with $k_x=k_y\equiv k_\perp=1,k_z=2$, involves $k_\perp^2k_z=2$ independent copies of $SU(3)$ irreps.    
\begin{figure}[!hbt]
      \includegraphics[width=0.5\textwidth]{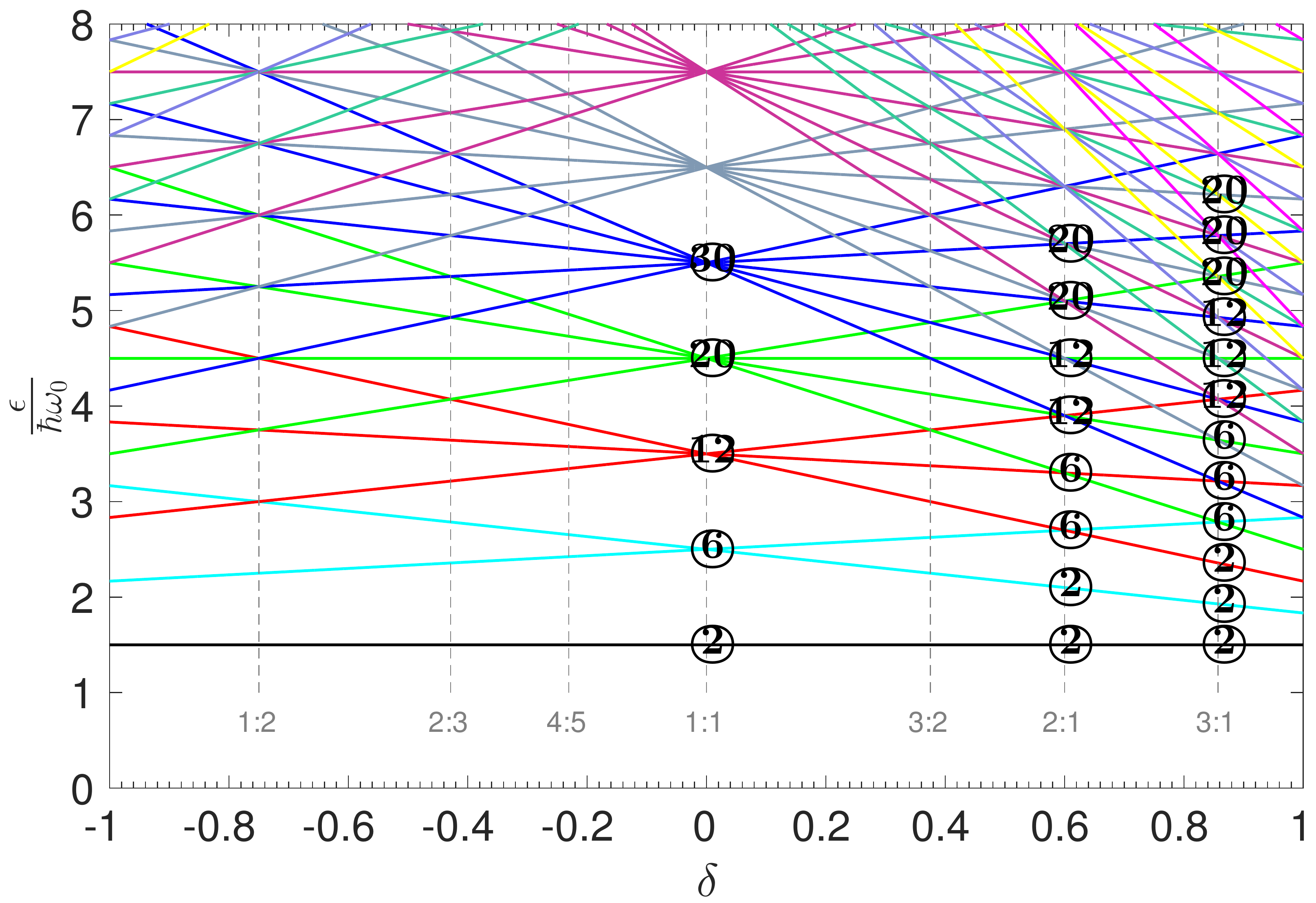}   
    \caption{(Color online) Level diagram of the 3D anisotropic 
    harmonic oscillator in axial symmetry versus the axial
    quadrupole deformation parameter $\delta$. Degeneracies 
    of the levels (taking into account the spin structure
    of nucleons) are indicated as well as deformations
    corresponding to rationally related frequencies $\omega_\perp:\omega_z$.   }
    \label{fig:HO}
  \end{figure}
As a consequence, the symmetries of the corresponding many-particle
wavefunction can be described by the irreps of two $SU(3)$ groups, suggesting that the shell structure of the superdeformed
HO is that of two smaller overlapping spherical HOs~\cite{naz92,fre95}. 
Moreover, the emerging superdeformed magic numbers 2, 4, 10, 16, 28, 40, \dots\ can be
expressed in terms of the spherical ones (2, 8, 20, \dots) either as the sum of two consecutive 
spherical magic numbers (2 = 2+0, 10 = 8+2, 28 = 20+8, \dots), or as repeating twice a spherical magic number 
(4 =2+2, 16 = 8+8, 40 = 20+20, \dots), see Fig.~\ref{fig:HO}. From these features, one can infer a susceptibility for nuclei in a superdeformed prolate state
to distribute their total mass among two spherical fragments, either in an asymmetric ($^{20}$Ne = $\alpha+^{16}$O, $^{56}$Ni = $^{16}$O$+^{40}$Ca, etc.) or a symmetric ($^{8}$Be = $\alpha+\alpha$, $^{32}$S = $^{16}$O$+^{16}$O, etc.) way. 
However, if deformation provides a necessary condition for nuclei to cluster in multiple 
spherical fragments via the occurrence of dynamical symmetries involving multiple independent 
copies of the $SU(3)$ irreps, it is not a sufficient one. As explained in Sec.~\ref{Sec:Loc}, 
other requirements must be met, e.g. a deep enough confining potential or a low enough mean 
density such that the spherical fragments do not overlap and dissolve by virtue of Pauli blocking effects.

\begin{figure}[h]
  \begin{center}
    \subfloat[]{
      \includegraphics[width=1\linewidth]{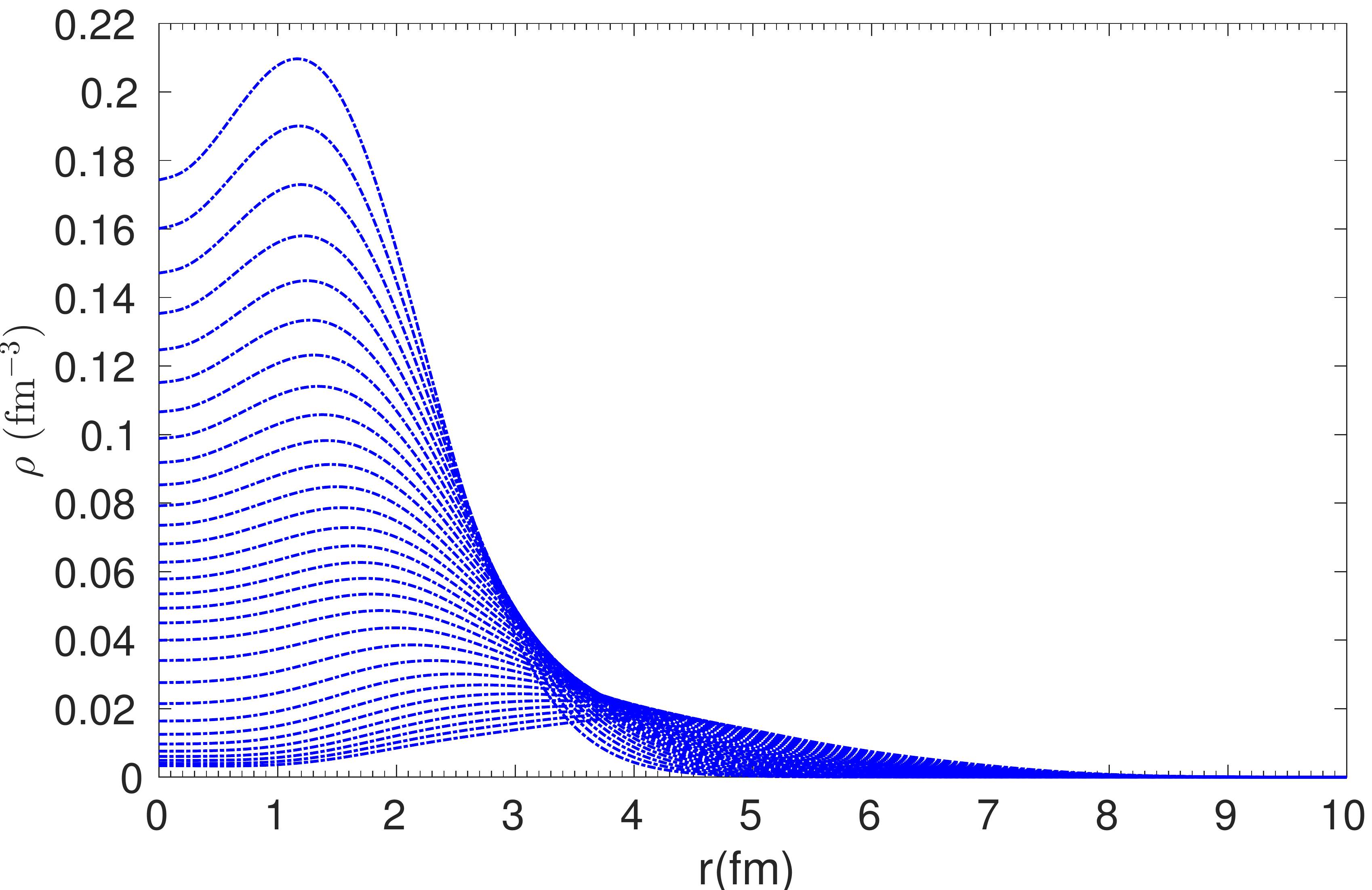}
      \label{sub:rmf}
                         }
    \\                     
    \subfloat[]{
      \includegraphics[width=1\linewidth]{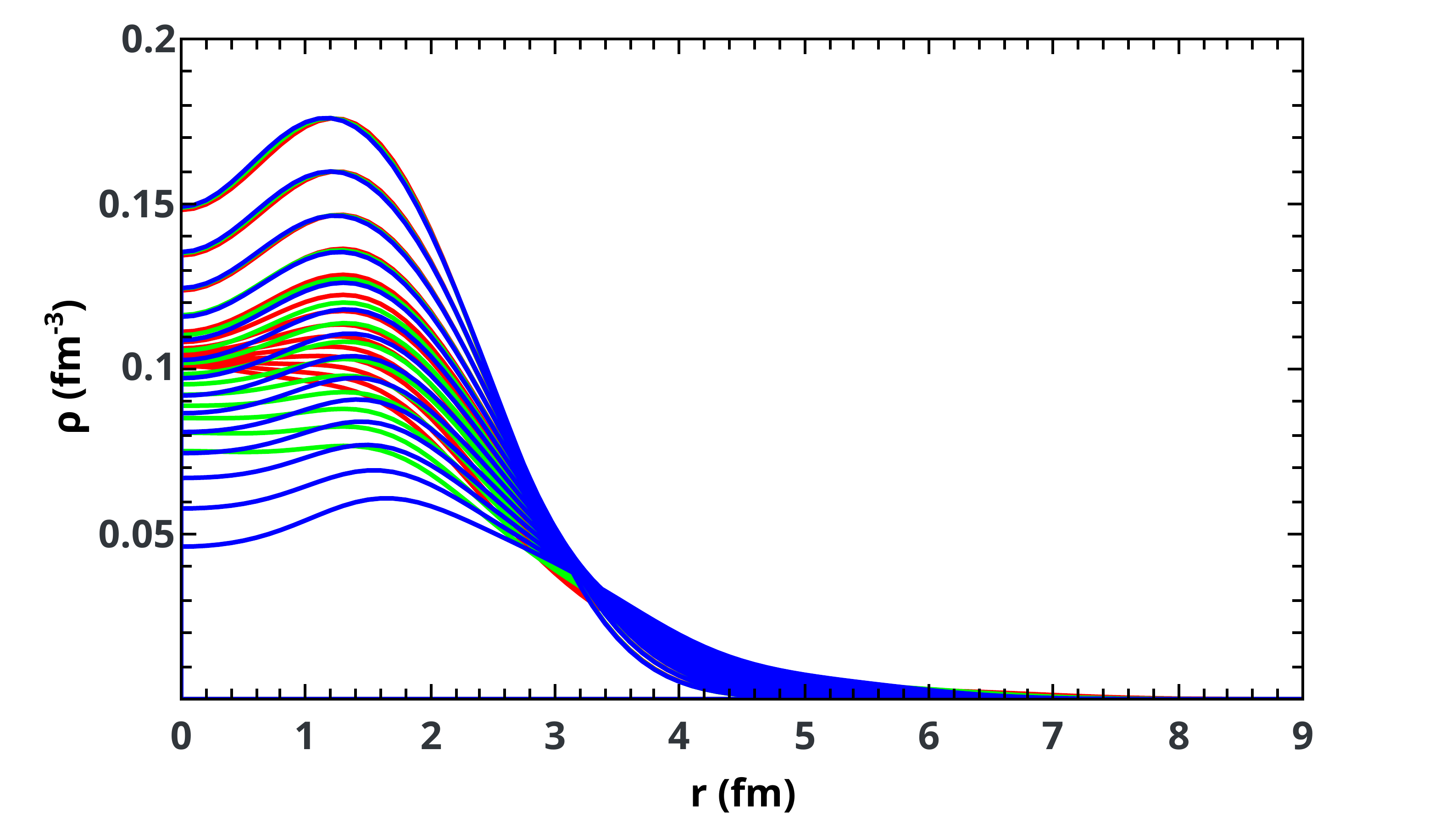}
      \label{sub:gogny}
                         }
    \caption{(Color online) $^{16}$O nucleon radial density for  
    r.m.s. radii constrained (a) from 2.4 to 5.3 fm within symmetry-restricted RMF calculations
    with the DDME2 parametrization and (b) from 2.5 to 3.8 fm within HFB calculations with the 
    Gogny D1S parametrization~\cite{ber91}. Relativistic calculations (a) are performed in a HO basis with 11 shells 
    and $\hbar\omega$ = 13 MeV. Nonrelativistic calculations (b) are also performed in a HO basis with 11 shells
    but with $\hbar\omega$ = 15 (red), 17 (green) and 19 (blue) MeV. }
    \label{fig:densrad}
  \end{center}
\end{figure}

\subsection{Role of the mean density in nuclear clustering}

We want to further investigate the impact of the mean density on the occurrence of clusters. 
Taking the specific case of $^{16}$O, this can be achieved by inflating isotropically the finite nucleus, i.e. by constraining its
r.m.s. radius while imposing a zero global quadrupole mass moment $Q_{20}$~\cite{gir13,ebr14}, 
such that the density continuously decreases. 

Such a program is worked out using both the CEDF and Gogny EDF approaches, where the corresponding constrained mean field 
equations are solved in a HO basis with 11 major shells. A more careful analysis of the impact of the HO basis parameters, 
e.g. the frequency $\omega$, must be undertaken when addressing such exotic dilute configurations. Indeed, some values of 
$\hbar\omega$ may lead to unphysical lower energy configurations where a part of nucleons remains tightly packed at the center
of the nucleus while the remaining nucleons are sparsely distributed around this dense core. We retain values of $\hbar\omega$ 
that minimize the energy of the system and at the same time lead to a regular decrease of the density.
These features are illustrated in Fig.~\ref{fig:densrad}.
The upper panel (a) displays the radial density of $^{16}$O for several constrained radii in the relativistic case.
The parameter of the HO basis yielding the lowest binding energy for these constrained configurations and a regular 
decrease of the density is found to be $\hbar\omega = 13$ MeV.
The lower panel (b) shows the density profile obtained in the non-relativistic case, where the colors distinguish between several values of $\hbar\omega$. For $\hbar\omega < 19$ MeV, the constraint on the radius, taken  between 2.4 and 3.8 fm, does not lead to a regular decrease of the density at the center of the nucleus. 
As a matter of fact, the nucleus increases its radius by expanding a low-density nucleon cloud surrounding a quasi constant-density nucleon core (red and green curves). These features do not sound appropriate. A regular decrease of the central density as the radius is constrained to higher values is only obtained for $\hbar\omega \sim 19$ MeV (blue curves) within Gogny D1S calculations, which is the adopted value in the non-relativistic case (as far as spherical configurations are concerned). 
\begin{figure}[h]
  \begin{center}
    \subfloat[]{
      \includegraphics[width=1\linewidth]{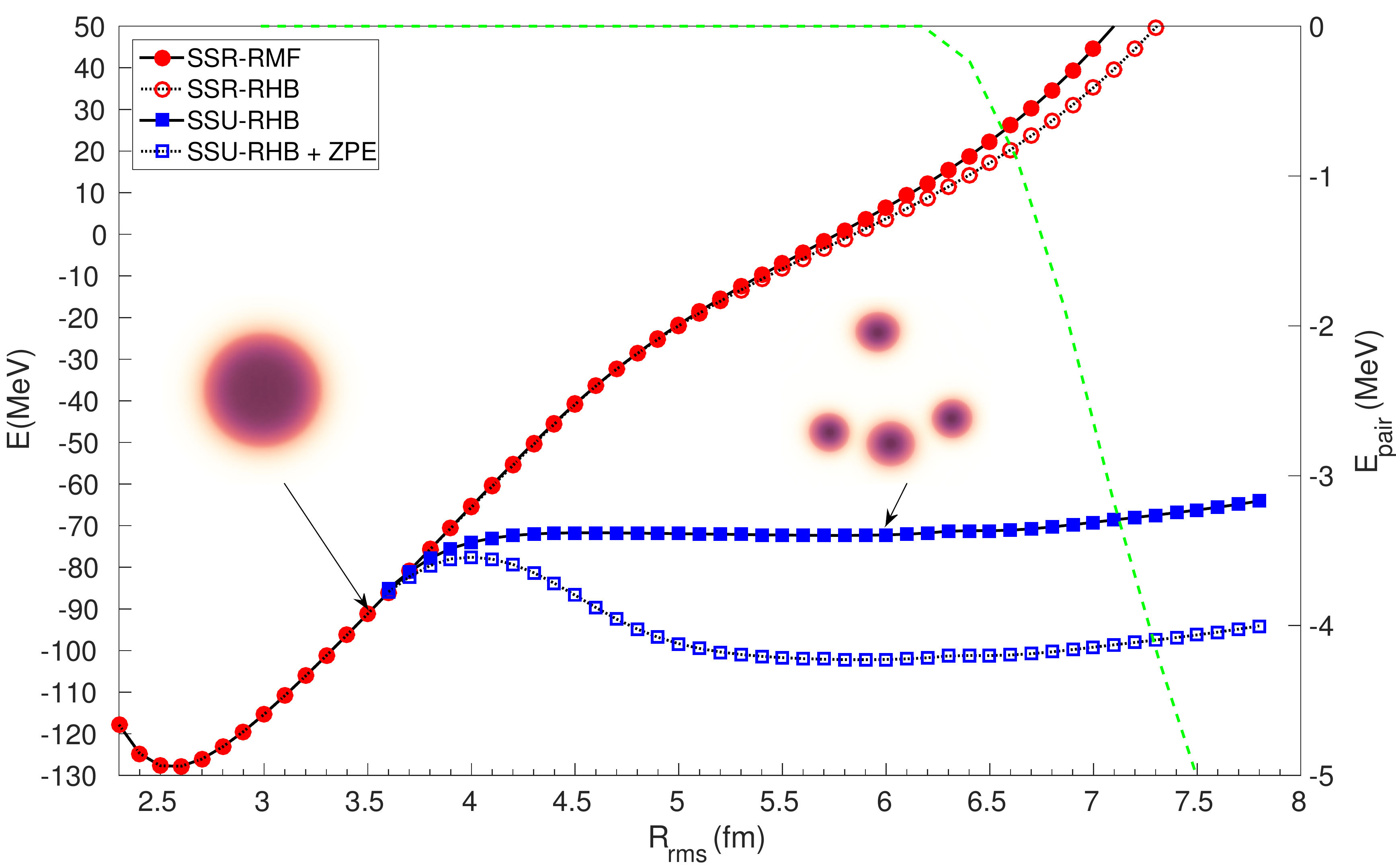}
      \label{sub:Ermf}
                         }
    \\                     
    \subfloat[]{
      \includegraphics[width=1\linewidth]{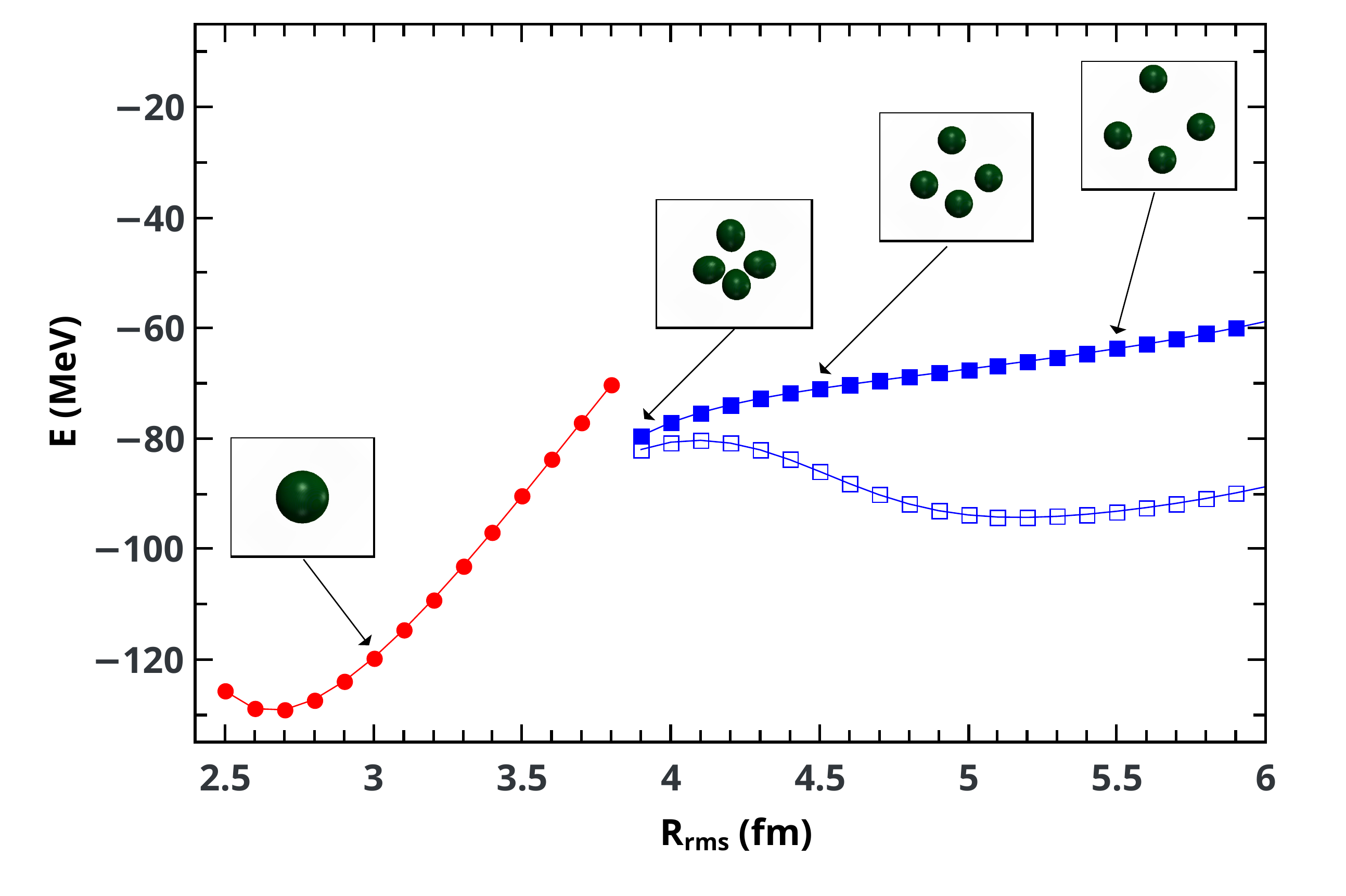}
      \label{sub:Egogny}
                         }
    \caption{(Color online) Self-consistent binding energy of $^{16}$O computed at the SR level of (a) CEDF with
    the DD-ME2 parametrization and (b) EDF with the Gogny D1S parametrization vs the constrained r.m.s. radius.
    In (a) curves with red circle symbols correspond to spherical configurations and were obtained in a 
    HO basis with 11 shells and $\hbar\omega$ = 13 MeV, while the blue square symbol curves
    correspond to a tetrahedral arrangement and were obtained in a HO basis with 11 shells and $\hbar\omega$ = 11 MeV.
    Likewise, in (b) red (circle symbols) and blue (square symbols) correspond to spherical and tetrahedral configuration 
    respectively and were obtained in a HO basis with 11 shells and $\hbar\omega$ = 19 and 12 MeV respectively. In both cases,
    the blue open square symbol curve corrects the mean field energy with the zero-point energy contribution, and the inserts 
    display $^{16}$O intrinsic density at the corresponding constrained radii. 
    See text for detailed explanations. }
    \label{fig:QPT}
  \end{center}
\end{figure}

Setting the HO basis parameters to their relevant values, $^{16}$O binding energy is computed as a function of the constrained
r.m.s. radius within both the CEDF and Gogny EDF approaches and displayed in Fig.~\ref{fig:QPT}.   
The relativistic case (Fig.~\ref{fig:QPT} (a)) involves several type of calculations. The curve with red filled circle markers corresponds to a SR-CEDF calculation
where we enforce spherical symmetry (i.e. no spatial spontaneous symmetry breaking (SSB) can occur) as well as the global U(1) invariance (i.e. no pairing correlations
can develop). We refer to this case as spatial symmetry-restricted relativistic mean field (SSR-RMF). 
Relaxing the enforcement of U(1) symmetry, i.e. still restricting the spatial symmetry
to the spherical one, but letting the system free to break the U(1) invariance signalling the development of pairing
correlations, yields the curve with red open circle markers : this is the spatial symmetry-restricted relativistic Hartree 
Bogoliubov (SSR-RHB) case. For this type of calculation, the green dashed line displays the corresponding pairing energy of the system.
Finaly, relaxing all the symmetry restrictions, both spatial and internal (with however the constraint $\beta_{20} = 0$ ensuring an istropic inflation of the nucleus) yields the curves with blue square markers (filled and open markers). This case is referred to as spatial symmetry-unrestricted relativistic Hartree Bogoliubov (SSU-RHB). In the curve with the square open symbols, the zero-point energy, computed as in ref.~\cite{gir13}, is substracted from the SSU-RHB energy.

\begin{figure}[!hbt]
      \includegraphics[width=1\linewidth]{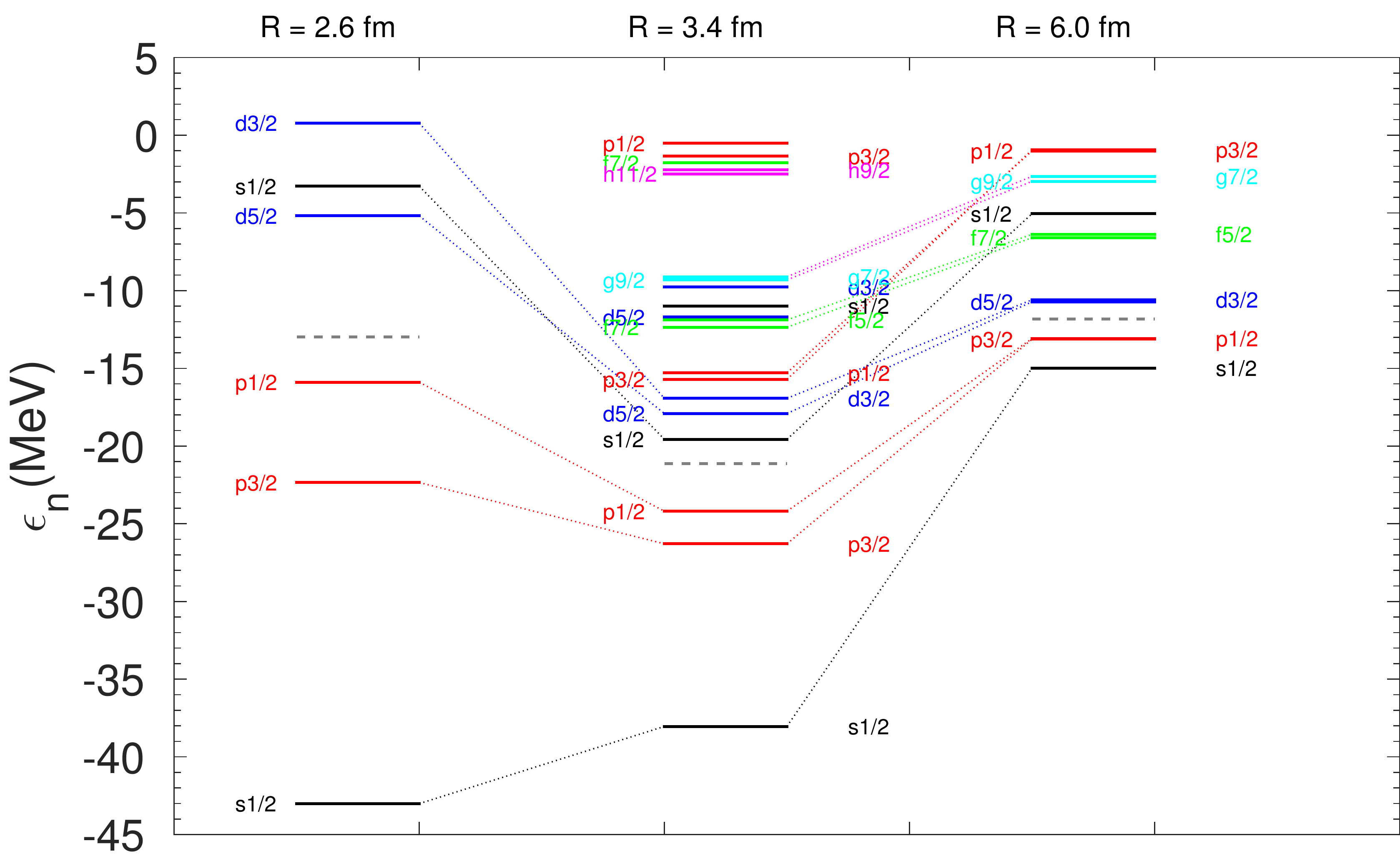}   
    \caption{(Color online) $^{16}$O neutron single-particle levels associated to
    SSR-RMF calculations with the DD-ME2 parametrization constrained at $R = 2.6, 3.4, 6.0$ fm.}
    \label{fig:RMFall}
  \end{figure}

\begin{figure}[h]
  \begin{center}
    \subfloat[]{
            \includegraphics[width=1\linewidth]{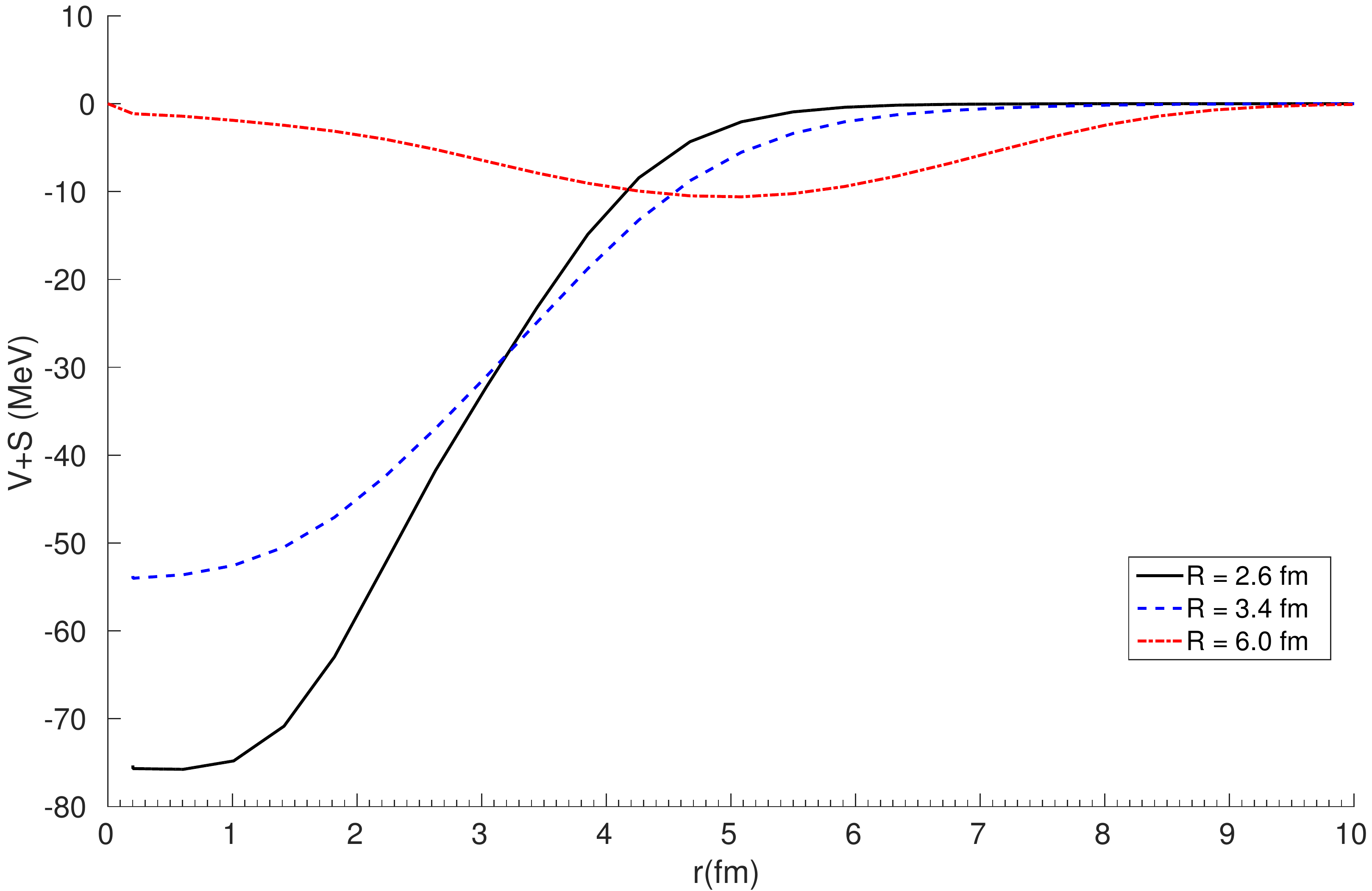}
      \label{sub:vps}
                         }
                         \\
    \subfloat[]{
       \includegraphics[width=1\linewidth]{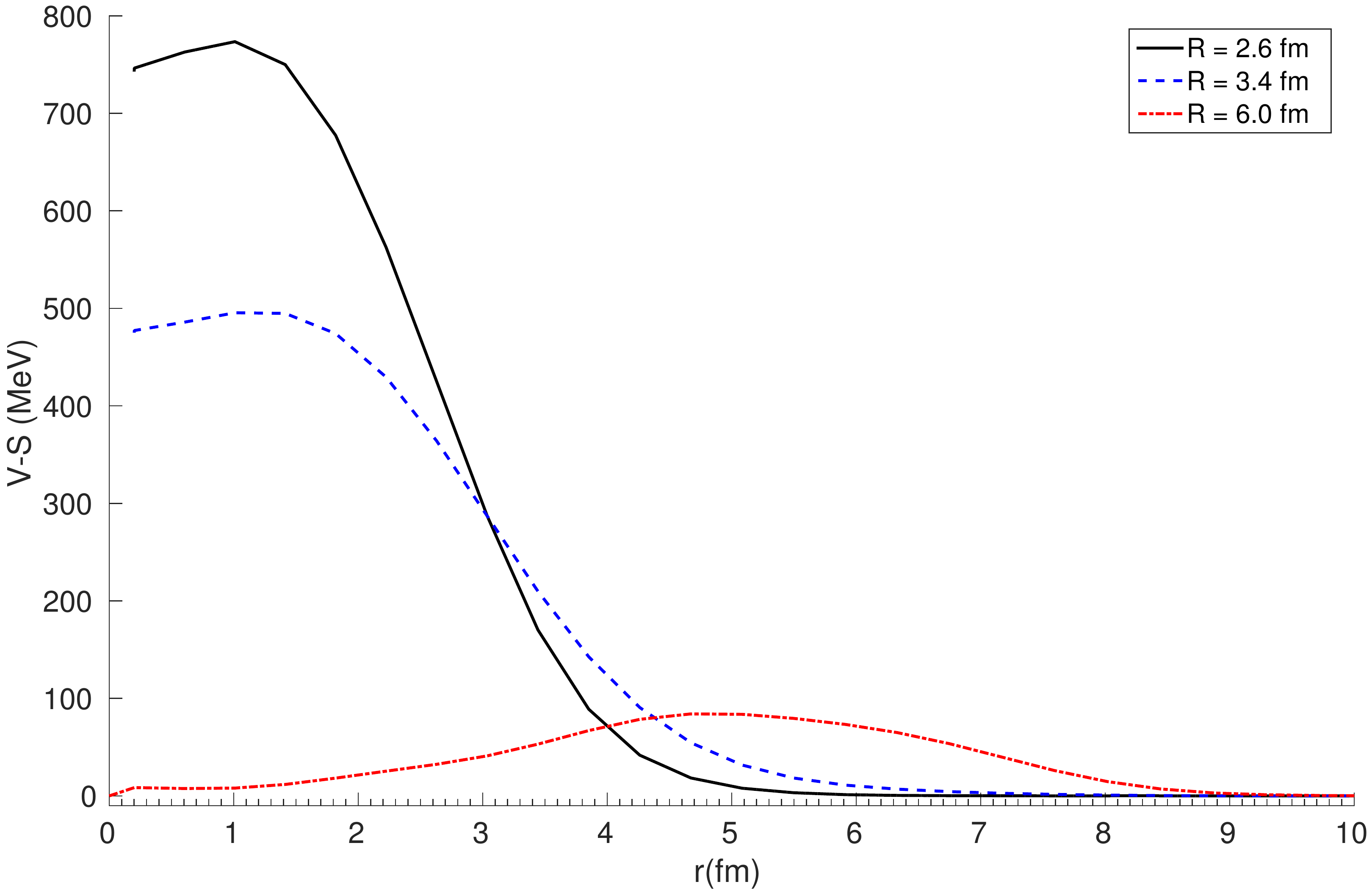}
      \label{sub:vms}
                         }
    \caption{(Color online) Radial evolution of combinations (a : $V+S$, b: $V-S$) of the 
    scalar ($S$) and time-like ($V$) nucleon self-energies in $^{16}$O for radii constrained
    at $R = 2.6, 3.4, 6.0$ fm.}
    \label{fig:vs}
  \end{center}
\end{figure}

Let us first analyse the SSR-RMF case. Fig.~\ref{fig:RMFall} displays the neutron single-particle (sp)
levels for three different constrained radii, namely the radius of the equilibrium configuration $R = 2.6$ fm,
a radius belonging to the interval where the three type of calculations (SSR-RMF, SSR-RHB and SSU-RHB) yield 
the same result $R = 3.4$ fm, and an extreme radius of $R = 6.0$ fm. 
Diluting $^{16}$O causes a drastic reduction of the valence neutron gap 
from 10.71 MeV at $R = 2.60$ fm  to 4.63 MeV at $R = 3.40$ fm and 2.36 MeV at $R = 6.0$ fm. The sp
spectrum gets shrinked and all spin-orbit partners eventually become degenerate. These features can be understood 
by looking at the radial dependence of two combinations of the scalar and time-like nucleon self-energies S and V (Fig.~\ref{fig:vs}).
The combination V+S defines the mean potential where independent nucleons evolve in the mean field picture. From 
a typical depth of $-75$ MeV at the equilibrium configuration, the confining potential become shallower as the
constrained radius increases until reaching $-10$ MeV at $R = 6.0$ fm. Likewise, the other combination V-S, whose derivative
(with a prefactor $1/M^2$ and M the nucleon mass) governs the spin-orbit splitting, gets weaker as the radius increases, restoring
the spin $SU(2)$ symmetry of the Dirac Hamiltonian, and therefore causing spin-orbit partners to be degenerate.

\begin{figure}[!hbt]
      \includegraphics[width=1\linewidth]{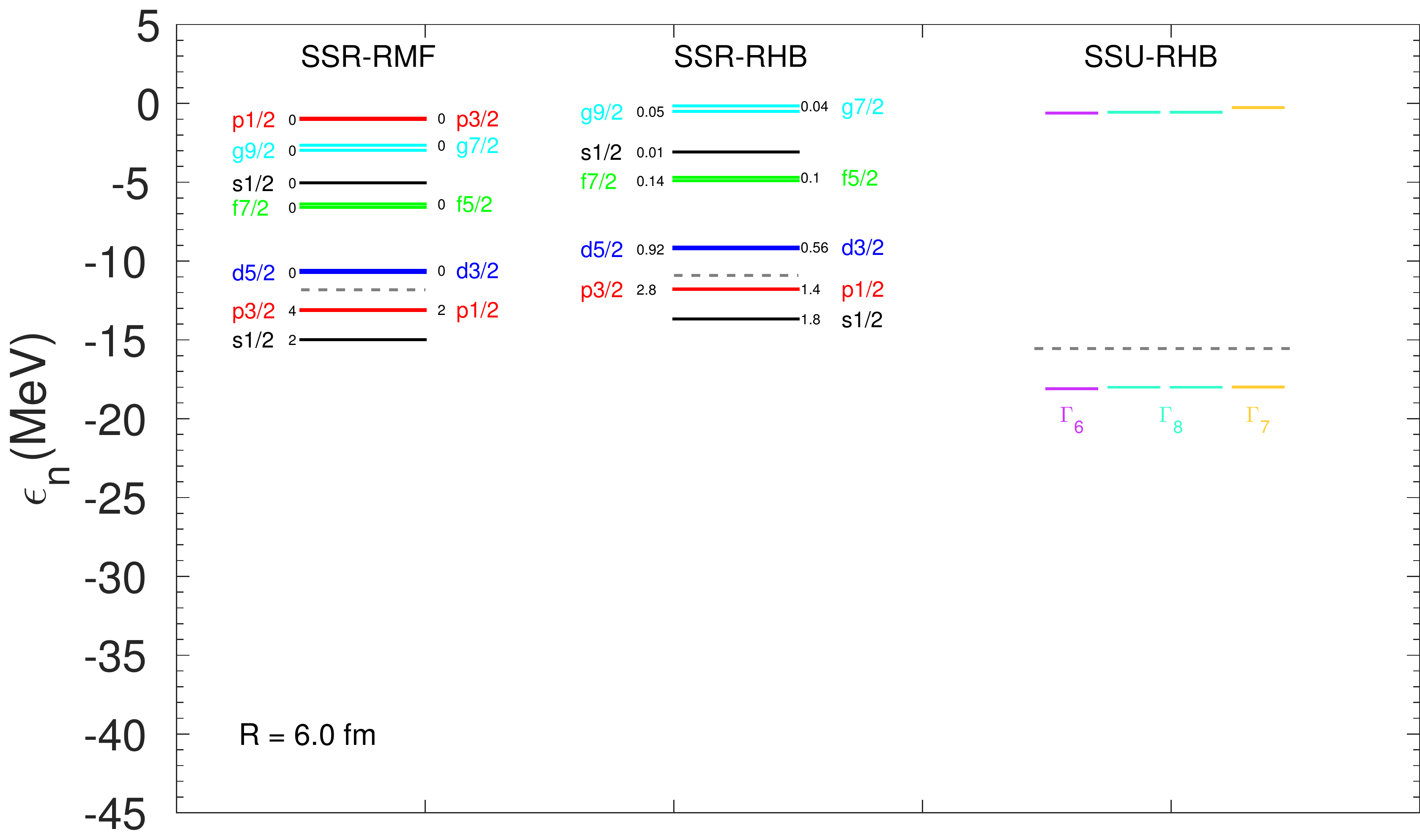}   
    \caption{(Color online) $^{16}$O neutron canonical single-particle levels computed 
    at $R = 6.0$ fm with the DD-ME2 parametrization in the SSR-RMF, SSR-RHB and SSU-RHB cases.}
    \label{fig:spectreall}
  \end{figure}

Such a reduction of the Fermi gap opens the room for the development of non-dynamical 
correlations. Indeed, $^{16}$O become a near degenerate system, i.e. excited particle-hole (ph)
configurations have energies close to the fundamental one, such that the system will rearrange
itself in a non perturbative way to lift the (near) degeneracies. A possible strategy consists in 
developing pairing correlations that can be accounted for, at the SR level, through the spontaneous breaking
of the global U(1) group associated to the conservation of nucleon number. 
From the corresponding SSR-RHB calculations displayed
Fig.~\ref{fig:QPT}, the normal to superfluid QPT occurs at $R = 3.8$ fm, even if a sensible effect on the binding 
energy has to wait for radii greater than $R = 5.5$ fm. Such a QPT translates into the opening of a gap in 
the quasi-particle (qp) spectrum of $^{16}$O and the strength of the corresponding correlations is measured by the pairing energy. At $R = 6.0$ fm, the gap in the neutron qp spectrum jumps from 2.2 MeV in the SSR-RMF case to 4.1 MeV in the SSR-RHB one, with a pairing energy of $\sim$ -8 MeV.
%
As one can see from the small impacts on $^{16}$O binding energy (Fig.~\ref{fig:QPT}), 
rearranging itself by developing pairing correlations seem rather ineffective because of   
the large energy splitting between the d and f levels that hinders the scattering of Cooper pairs, 
as shown by the occupation numbers of the canonical neutron sp spectrum in Fig.~\ref{fig:spectreall}.
A more effective strategy to lift the degeneracies is to develop angular correlations. From the SSU-RHB calculations,
a Mott-like QPT is observed at a critical radius $R_c = 3.7$ fm, that is a mean density
$\rho_\text{Mott}/\rho_0 = (R_\text{eq}/R_c)^3 \sim 0.35$, or $\rho_\text{Mott}\sim\rho_0/3$, in relatively good agreement with the estimation
based on dimensionless ratios of characteristic variables of the problem in Sec.~\ref{Sec:Loc}.   
The transition occurs between a phase where nucleons are delocalized in a dilute spherical volume
(the maximal value of the nucleon density is 0.046 fm$^{-3}$ at $R = 3.40$ fm) and a phase where nucleons are localized in 
four alpha-like degrees of freedom which recover a density equals to the saturation one and are arranged 
according to a tetrahedral configuration. For such a tetrahedral configuration, no pairing correlations develop.

The robustness of these results can be tested by performing similar calculations with the non-relativistic
Gogny EDF. The corresponding binding energy versus the constrained r.m.s. radius is displayed in Fig.~\ref{fig:QPT} (b) and
shows remarkable similarities with the relativistic case (Fig.~\ref{fig:QPT} (a)). Here the transition happens slightly later, 
at $R_c = 3.9$ fm, that is a mean density
$\rho_\text{Mott}/\rho_0 = (R_\text{eq}/R_c)^3 \sim 0.33$, i.e. again $\rho_\text{Mott}\sim\rho_0/3$.  
In the non-relativistic calculations, a break is observed between the curve related to the spherical configurations (in red) and those for the tetrahedral configurations (in blue). The discontinuity occurs at the radius beyond which the spherical density evolves into a four alpha-like configuration and stems from different values for the optimal $\hbar\omega$ in the spherical configuration case ($\hbar\omega = 19$ MeV) and in the tetrahedral one ($\hbar\omega = 12$ MeV). For the latter, the nucleus increases its radius by placing the alphas further apart, as illustrated in the inserts.


\subsection{Analysis of non-axial octupolar correlations}

\begin{figure}[!hbt]
      \includegraphics[width=1\linewidth]{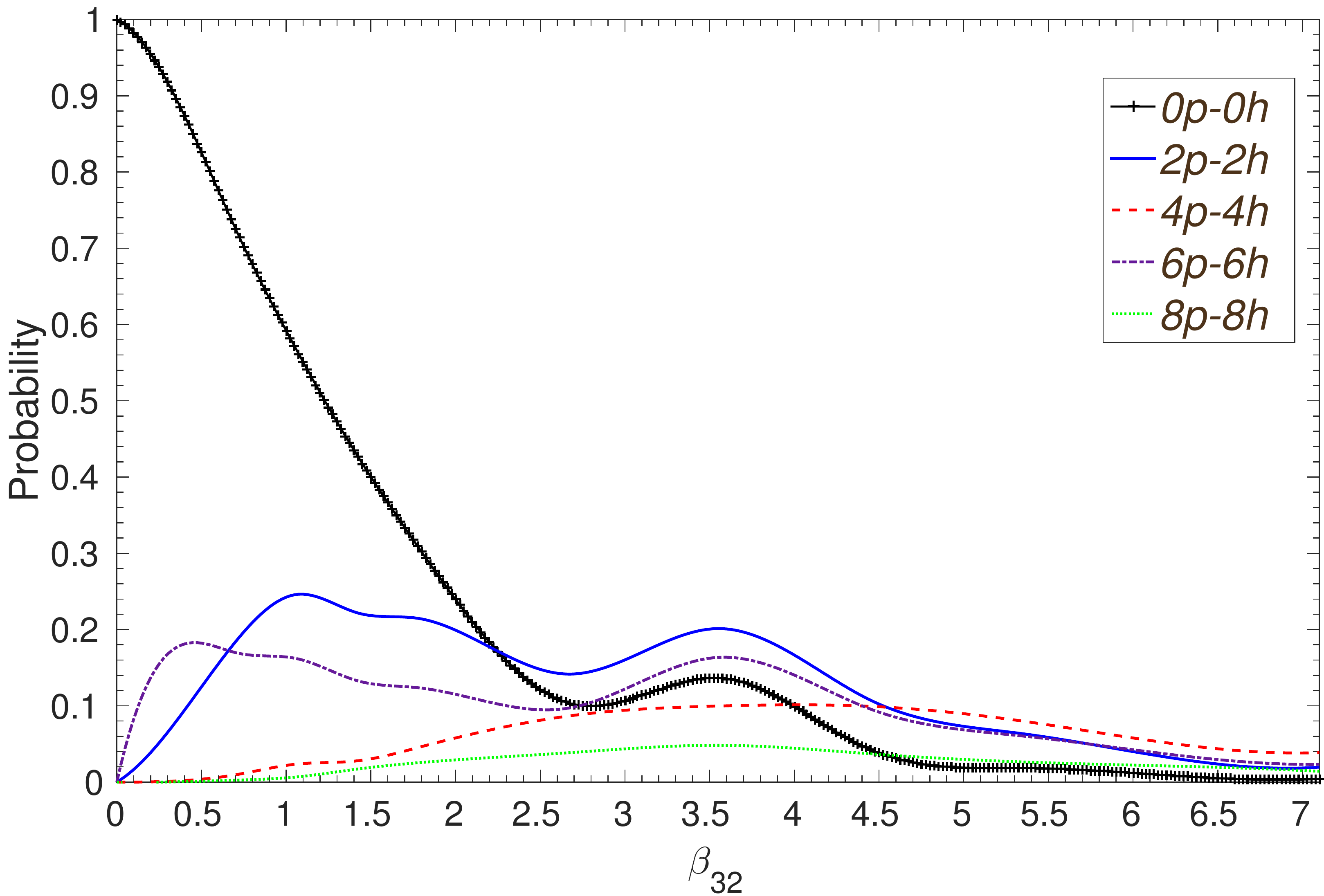}   
    \caption{(Color online) Probability to find n-particle-n-hole states belonging to the $O(3)$ irreps 
in the tetrahedral SD state as a function of the non-axial octupole deformation parameter $\beta_{32}$.}
    \label{fig:O16Proba}
  \end{figure}

\begin{figure}[!hbt]
      \includegraphics[width=1\linewidth]{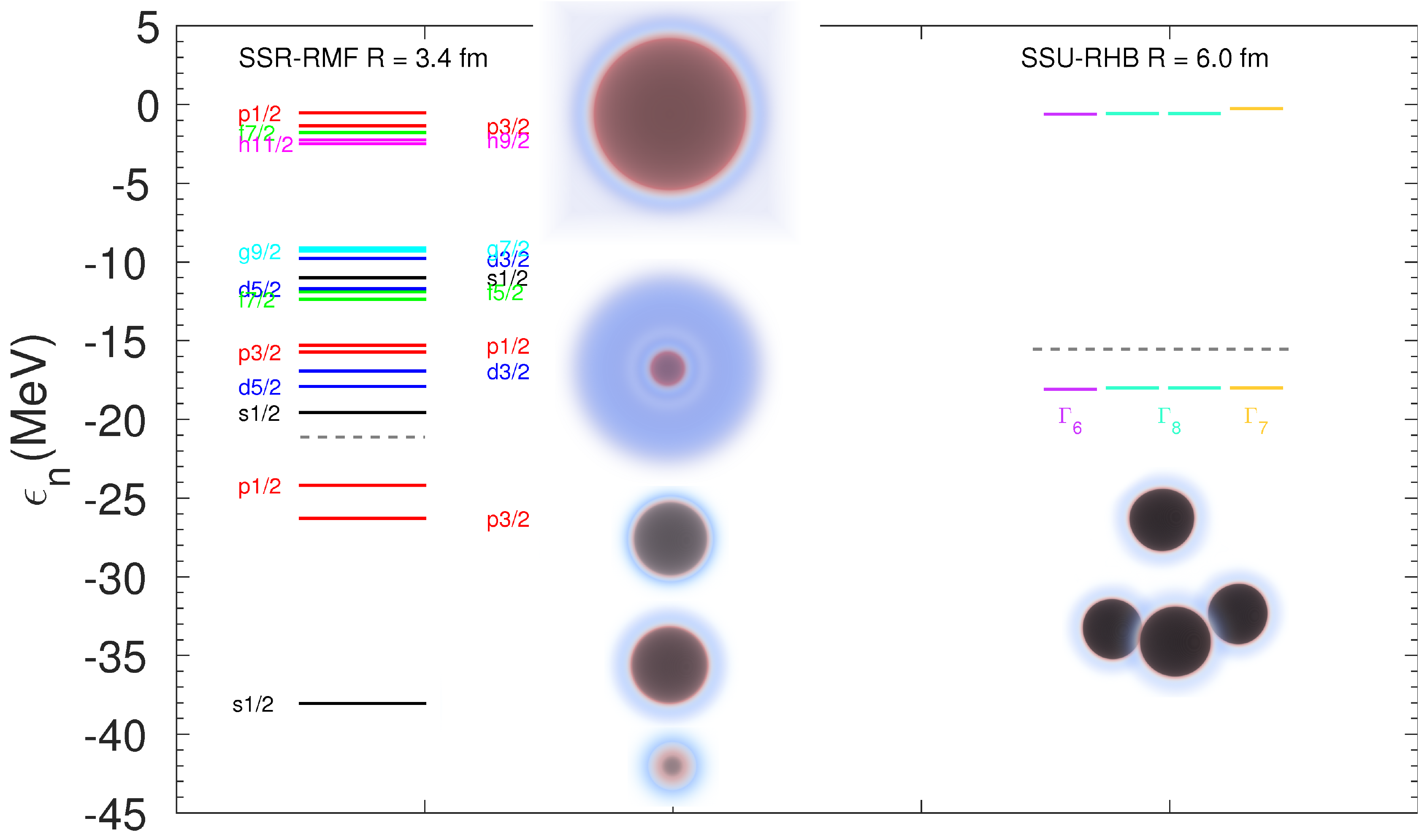}   
    \caption{(Color online) $^{16}$O neutron single-particle levels in the SSR-RMF case
    with $R = 3.4$ fm and in the SSU-RHB case with $R = 6.0$ fm computed with the DD-ME2
    parametrization. The inserts display the partial densities associated to the $1s1/2$,
    $1p3/2$, $1p1/2$, $2s1/2$ and $1f7/2$ orbitals from bottom to top in the SSR-RMF case, 
    and for the $\Gamma_6$, $\Gamma_7$, $\Gamma_8$ orbitals that share the same shape.}
    \label{fig:O16sphetetra}
  \end{figure}

\begin{figure}[h]
  \begin{center}
    \subfloat[]{
      \includegraphics[width=0.25\textwidth]{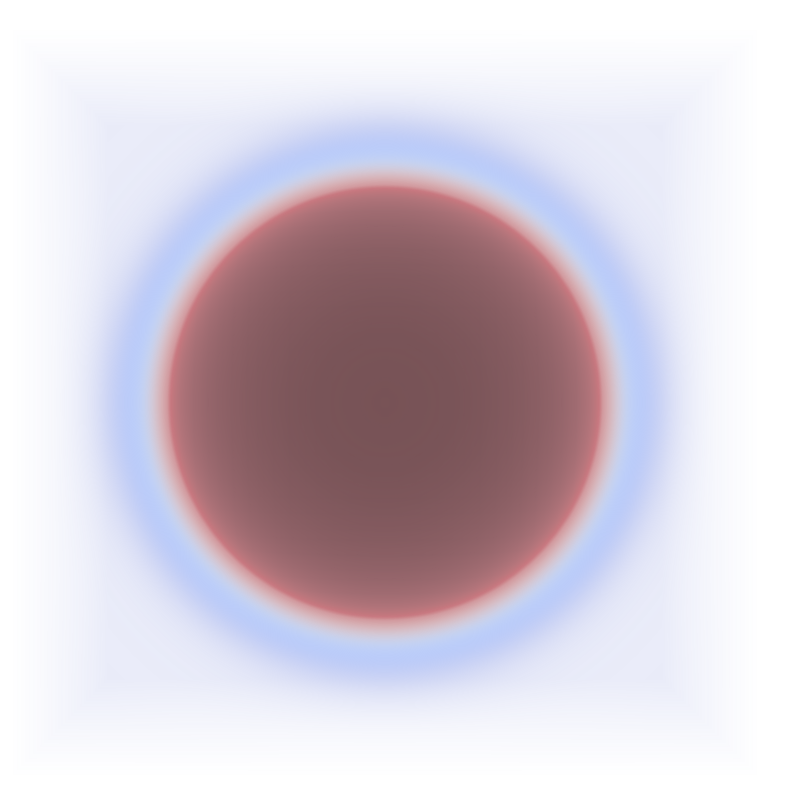}
      \label{sub:falone}
                         }
                         \\
    \subfloat[]{
      \includegraphics[width=1\linewidth]{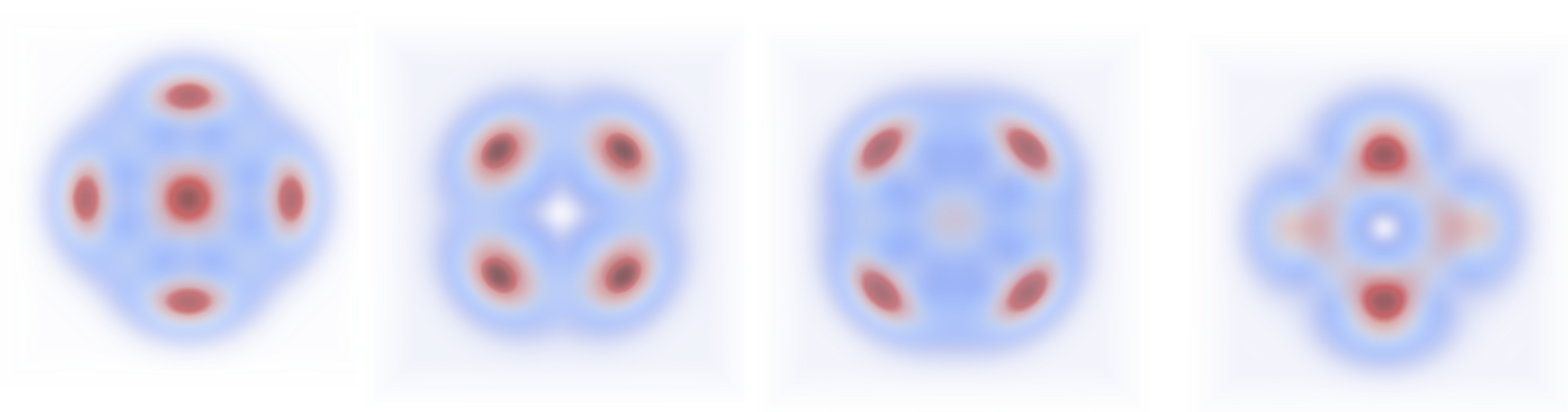}
      \label{sub:fdecomp}
                         }
    \caption{(Color online) $1f7/2$ partial density (a) and its decomposition
     on $m = 1/2,3/2,5/2,7/2$ (b). }
    \label{fig:f7}
  \end{center}
\end{figure}

The non-dynamical correlations grasped though the spatial SSB 
trigger the formation of alpha-clusters, allowing the dilute system to lower its energy by taking advantage of the nuclear cohesion. 
The corresponding neutron sp spectrum of the SSU-RHB calculation (Fig.~\ref{fig:spectreall}) displays a 
band-like structure, i.e. neutron sp levels assemble into two bunches of four
near degenerate orbitals separated by a huge gap of $17.4$ MeV.
We can further investigate the features of the tetrahedrally-deformed sp spectrum by considering
the SSB of the rotational group $O(3)$ down to the (double) point group $T_d$, 
whose character table is shown in Table~\ref{tab:char}.
\begin{table}[h]
\begin{tabular}{|c|c c c c c c c c|}
\hline  
           & $E$    & $\bar{E}$ & $8C_3$ & $8\bar{C}_3$ & $3C_2$          & $6S_4$       & $6\bar{S}_4$ & $6\sigma_d$ \\
           &        &           &        &              & $3\bar{C}_2$    &              &              & $6\bar{\sigma}_d$ \\
\hline
$\Gamma_1$ & 1       &   1      &  1     &     1        &       1         &       1      &      1       &    1      \\
$\Gamma_2$ & 1       &   1      &  1     &     1        &       1         &     - 1      &    - 1       &  - 1      \\
$\Gamma_3$ & 2       &   2      & -1     &   - 1        &       2         &       0      &      0       &    0      \\
$\Gamma_4$ & 3       &   3      &  0     &     0        &      -1         &       1      &      1       &  - 1      \\
$\Gamma_5$ & 3       &   3      &  0     &     0        &      -1         &      -1      &     -1       &    1      \\   
           &         &          &        &              &                 &              &              &                   \\
$\Gamma_6$ & 2       &  -2      &  1     &    -1        &       0         &   $\sqrt{2}$ & - $\sqrt{2}$ &    0      \\           
$\Gamma_7$ & 2       &  -2      &  1     &    -1        &       0         & - $\sqrt{2}$ &   $\sqrt{2}$ &    0      \\
$\Gamma_8$ & 4       &  -4      & -1     &     1        &       0         &   0          &   0          &    0      \\                      
\hline
\end{tabular}

\caption{Character table of the double Td group detailing its 8 classes and irreps. See~\cite{atk70} for further details.}
\label{tab:char} 
\end{table}
As spin-1/2 fermions, nucleon wavefunctions can be classified along the $\Gamma_6$, $\Gamma_7$ and $\Gamma_8$ irreps of 
the $Td$ group. The correlations grasped through the SSB of $O(3)$ down to $Td$ can be translated in the language of
ph excitations on top of a symmetry-preserving reference state by computing the overlaps 
$\braket{\Phi^\text{sphe}_i|\Phi^{tetra}_0(\beta_{32})}$
between the tetrahedral 
Slater determinant (SD)$\ket{\Phi^{tetra}_0(\beta_{32})}$ (the tetrahedral closed-shell configuration involving the $\Gamma_6$, $\Gamma_7$ and $\Gamma_8$ states) and spherical SDs $\ket{\Phi^\text{sphe}_i}$(both ground ($i=0$) 
and ph excitations ($i>0$) in a valence space spanning the 1s1/2 to the 1f5/2 states) for various values of the
order parameter $\beta_{32}$ (Fig.~\ref{fig:O16Proba}).
The contribution of the spherical closed-shell configuration (0p-0h state) drops 
rapidly as the tetrahedral deformation increases. Collective excitations, 
in particular 2p-2h and 6p-6h ones, quickly become dominant, meaning that the 
amplitude excitations from the p-shell to sd-shells do not describe the total
correlated wavefunction in a satisfactory manner, but one eventually needs to account
for pf-shell states as well as holes in the s state. It should be noted that the probabilities displayed in Fig.~\ref{fig:O16Proba} do not add up to one because of the too small valence space. 
The role played by the orbitals beyond the p- and sd-shells can be understood 
by comparing the shape of the spherical canonical orbitals with the tetrahedral ones (Fig.~\ref{fig:O16sphetetra}).
The four near degenerate tetrahedral orbitals $\Gamma_6$, $\Gamma_7$ and 
$\Gamma_8$ (doubly degenerate) share the same partial density that resembles
four alphas arranged in a tetrahedral configuration. To illustrate how spherically-symmetric
nucleonic shell combine in such tetrahedrally-deformed orbitals, 
let us focus on the $\Gamma_6$ level. From Table~\ref{tab:comp} that details the decompositions of 
the irreps of the full rotation group into irreps of the group Td, the irreps
compatible with $\Gamma_6$ are $D_\frac{1}{2}^+$, $D_\frac{7}{2}^+$, etc.  for the positive parity case and
$D_\frac{5}{2}^-$, $D_\frac{7}{2}^-$, etc.  for the negative parity one. 
\begin{table}[h]
\begin{tabular}{ c c| c| c c c| c c}
\hline  
\hline
                   &   &                                        &   &                       &   &                                       \\
 $D_0^+$           &   & $\Gamma_1$                             &   & $D_0^-$               &   & $ \Gamma_2$ \\
 $D_1^+$           &   & $\Gamma_4$                             &   & $D_1^-$               &   & $ \Gamma_5$ \\
 $D_2^+$           &   & $\Gamma_3+\Gamma_5$                    &   & $D_2^-$               &   & $ \Gamma_3+\Gamma_4$ \\
 $D_3^+$           &   & $\Gamma_2+\Gamma_4+\Gamma_5$           &   & $D_3^-$               &   & $\Gamma_1+\Gamma_4+\Gamma_5$ \\ 
 $D_4^+$           &   & $\Gamma_1+\Gamma_3+\Gamma_4+\Gamma_5$  &   & $D_4^-$               &   & $\Gamma_2+\Gamma_3+\Gamma_4+\Gamma_5$ \\
 $D_5^+$           &   & $\Gamma_3+2\Gamma_4+\Gamma_5$          &   & $D_5^-$               &   & $\Gamma_3+\Gamma_4+2\Gamma_5$ \\    
 $D_6^+$           &   & $\Gamma_1+\Gamma_2+\Gamma_3+\Gamma_4+2\Gamma_5$  &   & $D_5^-$               &   & $\Gamma1+\Gamma_2+\Gamma_3+2\Gamma_4+\Gamma_5$ \\ 
                   &   &                                        &   &                       &   &                                       \\ 
\hline
                   &   &                                        &   &                       &   &                                       \\
 $D_\frac{1}{2}^+$ &   & $\Gamma_6$                             &   & $D_\frac{1}{2}^-$     &   & $ \Gamma_7$ \\
 $D_\frac{3}{2}^+$ &   & $\Gamma_8$                             &   & $D_\frac{3}{2}^-$     &   & $ \Gamma_8$ \\
 $D_\frac{5}{2}^+$ &   & $\Gamma_7+\Gamma_8$                    &   & $D_\frac{5}{2}^-$     &   & $ \Gamma_6+\Gamma_8$ \\
 $D_\frac{7}{2}^+$ &   & $\Gamma_6+\Gamma_7+\Gamma_8$           &   & $D_\frac{7}{2}^-$     &   & $\Gamma_6+\Gamma_7+\Gamma_8$ \\ 
 $D_\frac{9}{2}^+$ &   & $\Gamma_6+2\Gamma_8$                   &   & $D_\frac{9}{2}^-$     &   & $\Gamma_7+2\Gamma_8$ \\         
 $D_\frac{11}{2}^+$ &   & $\Gamma_6+\Gamma_7+2\Gamma_8$         &   & $D_\frac{11}{2}^-$    &   & $\Gamma_6+\Gamma_7+2\Gamma_8$ \\ 
 $D_\frac{13}{2}^+$ &   & $\Gamma_6+2\Gamma_7+\Gamma_8$         &   & $D_\frac{13}{2}^-$    &   & $2\Gamma_6+\Gamma_7+2\Gamma_8$ \\       
                   &   &                                        &   &                       &   &                                       \\ 
\hline
\hline
\end{tabular}
\caption{Full rotation group compatibility table for the group $Td$~\cite{atk70}.}
\label{tab:comp} 
\end{table}
The corresponding lowest energy levels 
at $R = 3.4 $ fm can be read from Fig.~\ref{fig:O16sphetetra}: the occupied $1s1/2$ and unoccupied 
$2s1/2$, $1f7/2$ and $1f5/2$ orbitals (the p and d shells are not compatible with $\Gamma_6$). 
Superpositions of these (at least) four spherically-symmetric orbitals
are needed to yield a tetrahedrally-shaped $\Gamma_6$ orbital. The latter having a zero contribution at the center of 
the nucleus, one first need a mixture between the $1s1/2$ and $2s1/2$, which belong 
to the $\Gamma_6$ subspace, to cancel the density at the origin. The resulting density is 
still isotropically distributed in space. To localize the nucleons occupying the $\Gamma_6$ orbital
into alphas in a tetrahedral configuration, superposition with f states are also needed (since 
the p and d shells only involve the $\Gamma_7$ and $\Gamma_8$ irreps). Fig.~\ref{fig:f7} illustrates this
statement by splitting the $1f7/2$ partial density into its magnetic $m = 1/2,3/2,5/2,7/2$ degenerate components.    

The tetrahedral shape of the four $\Gamma_i$ ($i=6,7,8$) orbitals can be 
analysed by employing the language of quantum chemistry. These  
states resemble 'molecular' orbitals that have the structure
of a linear combination of localized 'atomic' orbitals, here 1s orbitals associated to 
the alpha-particle ground state.
In the LCAO-MO picture~\cite{mul67}, the molecular orbitals $\psi_i$ (i=1,2,3,4) read
\begin{equation}
\psi_i = \sum_{j=1}^4 f^i_j \phi_j,
\end{equation}    
where the $\phi_j$'s stand for the original Gaussian-type atomic orbitals.
The unknown coefficients $f^i_j$ follow from the resolution of an eigenvalue 
equation involving the norm and energy kernels $\mathcal{N}_{ij}$ and $\mathcal{H}_{ij}$
between the atomic orbitals. In the H\"uckel approximation, $\mathcal{N}_{ij} = 0, \forall i,j $  
while all $\mathcal{H}_{ij} = 0$ except when the ith and jth atomic sites are adjacent. 
Denoting $\epsilon \equiv \mathcal{H}_{ii} \forall i $ and $-\mu \equiv \mathcal{H}_{ij}$ for all 
the other ajacent i and j, the H\"uckel eigenvalue equation for 4 alphas located on the vertices
of a tetrahedron reads
\begin{equation}
\label{eq:Hu}
\begin{pmatrix}
\epsilon & -\mu & -\mu & -\mu  \\
 -\mu & \epsilon & -\mu & -\mu \\
 -\mu & -\mu & \epsilon & -\mu \\
 -\mu & -\mu & -\mu & \epsilon
\end{pmatrix} \begin{pmatrix}
f^i_1 \\
f^i_2 \\
f^i_3 \\
f^i_4
\end{pmatrix}
 = E_i \begin{pmatrix}
f^i_1 \\
f^i_2 \\
f^i_3 \\
f^i_4
\end{pmatrix}.
\end{equation}
The resolution of Eq.~\eqref{eq:Hu} yields the four LCAO-MO's:
\begin{equation}
\psi_1 = \frac{1}{2} \left(\phi_1 + \phi_2 + \phi_3 + \phi_4 \right), 
\end{equation} 
with energy $E_1 = \epsilon -3\mu$;
\begin{equation}
\psi_2 = \frac{1}{\sqrt{2}} \left(-\phi_1 + \phi_2 \right), 
\end{equation} 
with energy $E_2 = \epsilon + \mu$;
\begin{equation}
\psi_3 = \frac{1}{\sqrt{2}} \left(-\phi_1 + \phi_3 \right), 
\end{equation} 
with energy $E_3 = E_2$;
\begin{equation}
\psi_4 = \frac{1}{\sqrt{2}} \left(-\phi_1 + \phi_4 \right), 
\end{equation} 
with energy $E_4 = E_3 = E_2$.
Being degenerate, the three last orbitals are not uniquely defined. Any
linear combination involving a $3\times 3$ unitary transformation yields a triplet
of different but equivalent orbitals, e.g.
\begin{eqnarray}
\psi'_2 &=& \frac{1}{2} \left(\phi_1 - \phi_2 - \phi_3 + \phi_4 \right), \nonumber \\
\psi'_3 &=& \frac{1}{2} \left(\phi_1 + \phi_2 - \phi_3 - \phi_4 \right), \nonumber \\
\psi'_4 &=& \frac{1}{2} \left(-\phi_1 + \phi_2 - \phi_3 + \phi_4 \right). 
\end{eqnarray}
These results hence account for the structure of the symmetry-broken sp spectrum. 
The SSB of $O(3)$ down to the point group $Td$ translates into the clusterization
of the nucleon in four alphas arranged according to a tetrahedral configuration. The latter brings into
play four nearly degenerate (up to the value of $\mu$ for the first orbital ) molecular-like orbitals
that can be expressed as linear combinations of four 1s alpha states. At $R = 6.0$ fm 
the energies of the $\Gamma_i$ ($i=6,7,8$) $E_1 = -18.090$ MeV and $E_2 = E_3 = E_4 = -18.004$ MeV (see Fig.~\ref{fig:O16sphetetra})
lead to identify the energy of the 1s alpha state $\epsilon = -18.004$ MeV  as well as the non-diagonal energy kernel 
$\mu = 0.086$ MeV. Taking into account the $\sim 10$ MeV correction coming from the zero-point energy contribution 
(blue open square curve in Fig.~\ref{fig:QPT}), the energy of the 1s alpha state drops to $\sim -28$ MeV, in agreement 
with the binding energy of $^4$He. It should be noted that the energy gap of $17.4$ MeV between occupied and unoccupied states of the tetrahedrally deformed sp spectrum is of the order of
the lowest excitation energy of the alpha-particle (20.2 MeV), suggesting an exclusion property that acts among the nucleons sharing the same intrinsic state
when embedded in an alpha-cluster.

\section{Quartet quantum phase transition in infinite matter}
\label{Sec:Schuck}

So far, we have considered a QPT in $^{16}$O as a function of the density where the nucleus changes from a homogeneous mean-field density spontaneously into a tetraheadral configuration of four $\alpha$ particles. However, those cristalline structures, imposed by the mean-field, come  too high in energy. It is, for example, known that it is difficult to describe in detail the famous Hoyle state in $^{12}$C at 7.65 MeV in this way. However effects of the Pauli principle and the density are already well given in mean field theory.

We here want to study quartet condensation and the corresponding  QPT in infinite matter and make a link with the preceding mean-field study of $^{16}$O concerning the typical densities at which the QPT occurs in nuclear systems.
Quartet condensation is described following very closely the usual procedure of pairing with the BCS approach. For the latter the BCS equations can be written in the following way

\begin{equation}
  (e_{p_1} + e_{p_2})\kappa_{p_1p_2} + (1-n_{p_1} - n_{p_2}) \sum_{p'_1p'_2}v_{p_1p_2p'_1p'_2}\kappa_{p'_1p'_2} = 2\mu\kappa_{p_1p_2},
  \label{2body}
\end{equation}
with the occupation numbers given by

\begin{equation}
  n_k = \frac{1}{2}\bigg [ 1- \frac{e_k - \mu}{\sqrt{(e_k-\mu)^2 + \delta_k^2}}\bigg ],
    \label{occs}
\end{equation}
with the gap

\[ \Delta_k = g \kappa_{k\bar k}, \]
where $\bar k$ is the time reversed state of $k$ and we used as pairing force a delta interaction $g\delta({\bf r}_1-{\bf r}_2)$. Finite range forces can be treated accordingly.\\
In above equations $e_k$ are the kinetic energies, eventually with inclusion of a Hartree-Fock (HF) shift, and $\mu_i$ is the chemical potential. The indices $p$ include momenta and spin, $\kappa_{p_1p_2} = \langle c_{p_1}c_{p_2}\rangle$ is the pairing tensor, and $v_{p_1p_2p_3p_4}$ is the matrix element of the pairing force. Equations (\ref{2body}) and (\ref{occs}) are the BCS equations in their general form. Usually one considers the Cooper pairs at rest, what makes that the momenta of the two particles are opposite and one considers spin singulet pairing.

\begin{figure}
\includegraphics[width=1\linewidth]{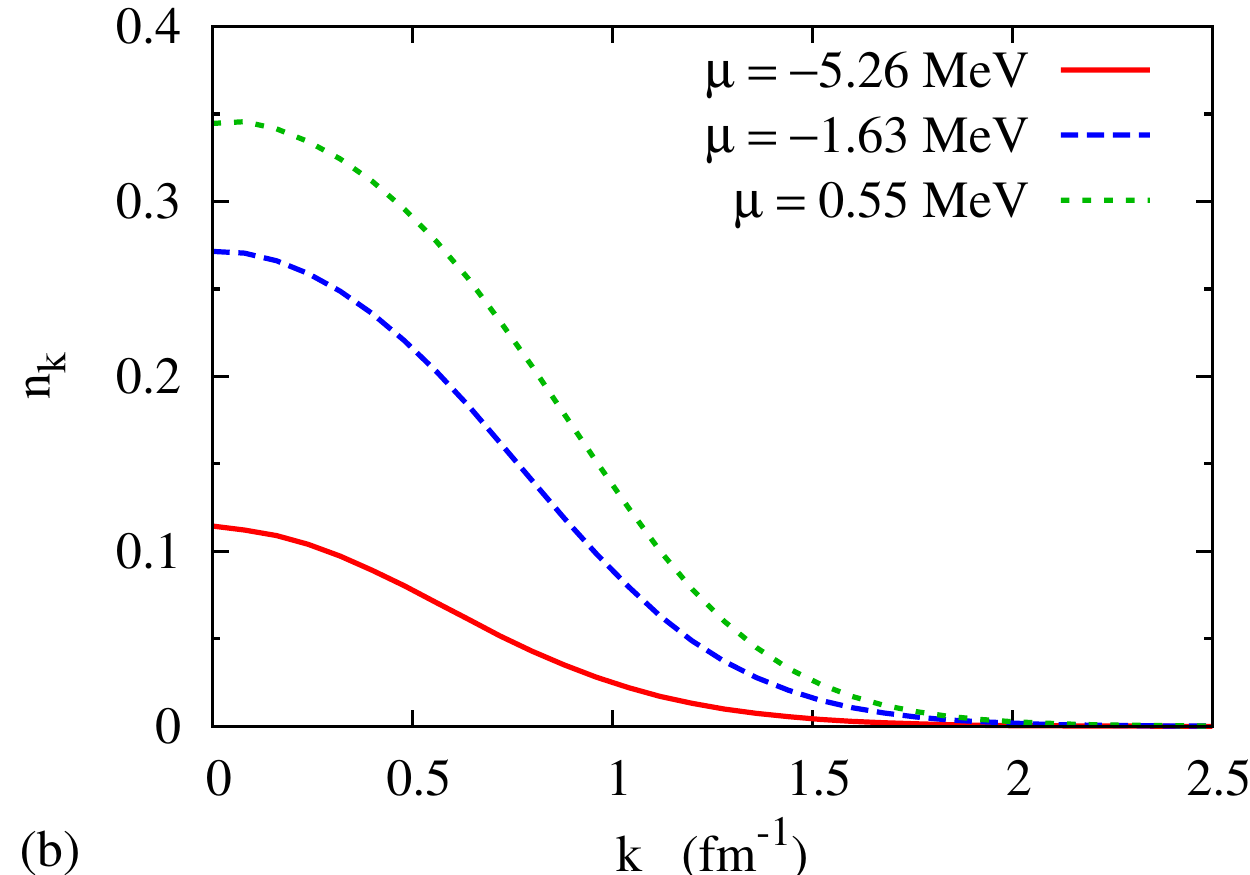}
\caption{\label{a-occs} Single nucleon occupation numers $n_k$ for different values of mu. The highest value before the calculation of the $\alpha$-order parameter breaks down is $\mu \sim 0.55$ MeV.}
\end{figure}

\begin{figure}[htbp]
\begin{center}
\includegraphics[width=1\linewidth]{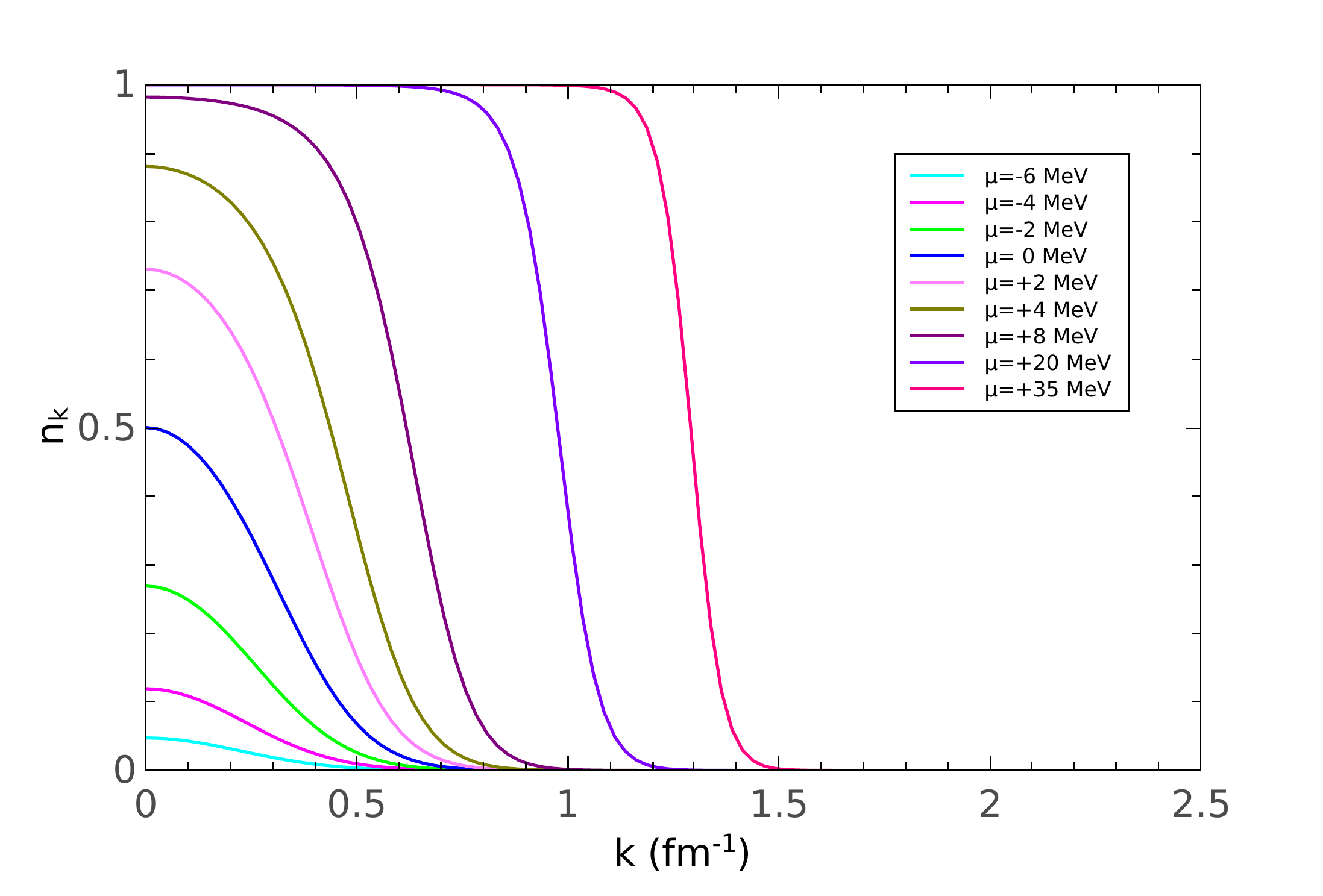}
\end{center}
\caption{Schematic (non-selfconsistent) view of BCS occupation numbers as the chemical potential varies from positive to negative (binding) values. }
\label{occs-BCS}
\end{figure}

For quartetting, one proceeds in a completely analogous way: one writes the in medium four body equation \cite{Sogo2}

\begin{equation}
  (e_1+e_2+e_3+e_4)\kappa_{1234} + \sum_{1'2'3'4'}V_{1234;1'2'3'4'}\kappa_{1'2'3'4'}=4\mu\kappa_{1234},
  \label{4body}
\end{equation}
with
\begin{eqnarray}
  V_{1234;1'2'3'4'} = &(&1-n_1-n_2)v_{121'2'}\delta_{33'}\delta_{44'}\nonumber\\
  &+& (1- n_1 - n_3) v_{131'3'} +{\rm permutations},\nonumber \\
\end{eqnarray}
where we used an obvious short hand notation. In the case of quartetting the expressions for the occupation numbers $n_k$ are quite a bit more complicated with respect to the pairing case and we refer the reader to the literature \cite{Sogo2}. To ease the numerical solution of the quartet equation, in Ref. \cite{Sogo2}, 
the four nucleon order parameter was approximated by a mean-field ansatz projected to good total center of mass momentum $K=0$ in the following way

\begin{equation}
\langle c^+_{{\bf k}_1}c^+_{{\bf k}_2}c^+_{{\bf k}_3}c^+_{{\bf k}_4}\rangle = 
\delta({\bf k}_1 + {\bf k}_2+{\bf k}_3+{\bf k}_4)\varphi({\bf k}_1)\varphi({\bf k}_2)\varphi({\bf k}_3)\varphi({\bf k}_4),
\end{equation}
where $c^+_{\bf k}$ creates a nucleon with momentum ${\bf k}$ (obvious spin-isospin indices are suppressed as well as the total scalar spin-isospin part of the wave function) and $\varphi({\bf k})$ is a 0S single particle wave function in momentum space. The selfconsistent equation for the order parameter then boils down to a nonlinear equation for $\varphi({\bf k})$ and it turned out that this approximation reproduces very well a full solution of the in medium four body equation \cite{Sogo1}. The point now is that this order parameter only exists below a critical density of $\sim \rho_0/5$~\cite{rop14} which is a value similar to what found in the preceding study for $^{16}$O. In infinite matter, this can then be qualified as a macroscopic QPT for quartets ($\alpha$ particles) with the density as control parameter.\\ 
This breakdown was studied with the calculation of the single nucleon occupation numbers $n_k$ in the $\alpha$-condensate as a function of the chemical potential $\mu$. We see in Fig.\ref{a-occs} that as $\mu$ increases, $n_k$ naturally  also increase. However, beyond $\mu \sim 0.55$ MeV where $n_{k=0} \sim 0.35$, the solution ceases to exist, that is the $\alpha$ order parameter has disappeared and the system has turned over into a standard nuclear superfluid very analogous to what we have seen happening in $^{16}$O. It should be pointed out that this behavior is in strong contrast to pairing, for instance deuteron pairing, where the density can be increased without breakdown of superfluidity, the decrease of the gap being only connected with the finite range of the pairing force. The corresponging $n_k$ steadily increase from negative to positve values of $\mu$ without interruption. 
Of course the Pauli principle forbids that $n_k$ overshoots the value of one (disregarding spin and iso-spin degeneracies) reaching the typical BCS-like behavior at nuclear saturation densities. This behavior is shown in Fig. \ref{occs-BCS} in a qualitative way. We see the strong difference with the behavior of $n_k$ in the quartet case. It should be noted that the distributions below and around $\mu 
\simeq 0$ should be compared with the ones of the quartet case. 

\begin{figure}[htbp]
\begin{center}
\includegraphics[scale=1.]{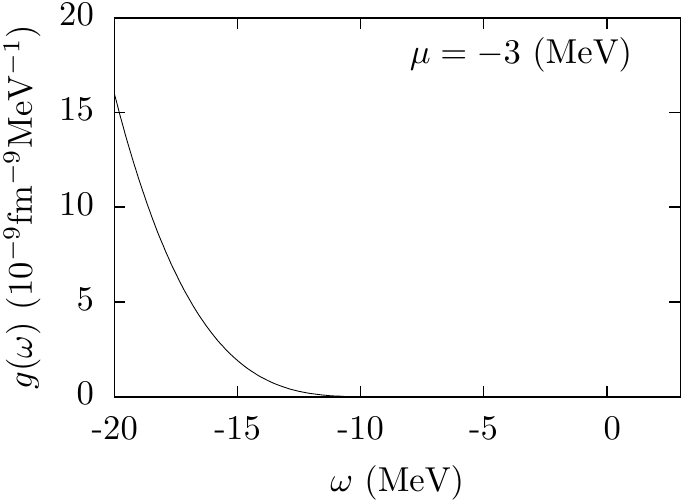}
\includegraphics[scale=1.]{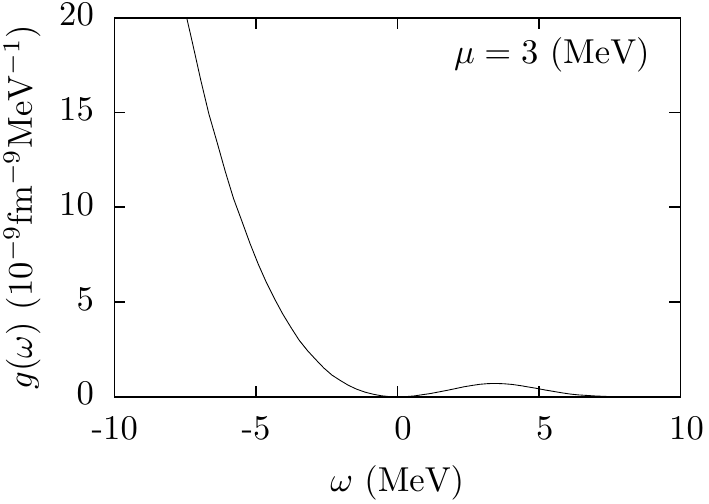}
\includegraphics[scale=1.]{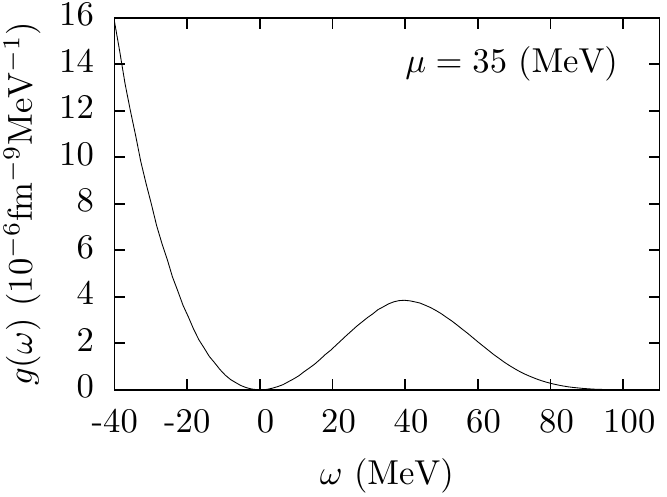}
\end{center}
\caption{\label{Sogoleveldensity} 3$h$-level density for negative (top)
  and positive (bottom) chemical potentials \cite{Sogo1}.}
\end{figure}

The reason for this breakdown has a simple physical interpretation. It seems clear that in a four-body problem the in medium four-body level density plays a dominant role. It is defined by \cite{Blin}

\begin{eqnarray}
g_4(\omega) = \sum_{k_1,k_2,k_3,k_4}\left[\bar f_1\bar f_2\bar f_3\bar f_4 - f_1f_2f_3f_4\right] \nonumber \\
\delta(\omega- e_1-e_2-e_3-e_4),
\end{eqnarray}
where $\bar f = 1-f$ and $f_i =f(e_i)$ is the Fermi-Dirac function equal to $\Theta(\mu - e_i)$ at zero temperature. The $e_i$ are the kinetic energies $p^2_i/(2m)$. One easily verifies that for positive chemical potential $\mu$, this four-body level density goes through zero at $\omega = 4\mu$. Where there is no level density at the Fermi surface no correlations can develop and, thus, the order parameter goes to zero very soon after $\mu$  has turned from negative values (binding) to positive ones (scattering). Actually for the calculation of the $n_k$ shown in Fig. \ref{a-occs}, one needs the three hole level density \cite{rmp,Sogo2}

\begin{equation}
g_{3h}(\omega) = \sum_{k_1k_2k_3} \left[\bar f_1\bar f_2 \bar f_3 + f_1f_2f_3\right]\delta(\omega
+ e_1 + e_2 + e_3),
\label{3h-dos}
\end{equation}
and for convenience we show that level density here for three values of the chemical potential, see Fig. \ref{Sogoleveldensity}. Actually all many-particle level densities besides the one for two particles, where the pair is at rest, go through zero at the Fermi level, see \cite{Blin} and below. 
For the case of negative $\mu$ the Fermi-Dirac functions at zero temperature are zero and then there is no qualitative difference for any of the multiparticle level densities, since the phase space factors reduce to unity. 

Actually the disappearance of the $\alpha$-particle condensate does not coincide with the appearance of an uncorrelated Fermi gas as we here supposed. Rather, as already mentioned,  there will appear a superfluid Fermi gas. However, a superfluid Fermi gas shows a gap at the Fermi level (chemical potential) and thus the level densities are even more suppressed around the Fermi level than without pairing.

The critical density coincides with the Mott density at zero temperature \cite{rop14}. Actually the critical density is just the one where the alphas start to overlap with their tails to some appreciable extent (see, e.g., the two alphas in $^8$Be \cite{wir00}) and, thus, the Pauli principle becomes active. For the pairing case, for two particles at rest with their c.o.m., one verifies that the two-body level density is finite at $\omega = 2\mu$, this being the reason why pairing also exists for positive $\mu$, or at high densities. For example the two-body level density for two particles below two times  the chemical potential is given by

\begin{equation}
  g(\omega)_{2p} = \sum_{k_1k_2}\Theta(\mu-e_{k_1})\Theta(\mu - e_{k_2})\delta(\omega - e_{k_1} -e_{k_2}).
  \label{leveldensity-2p}
  \end{equation}
For the particle pair at rest $k_1=k_2$, one easily verifies that for $\omega=2\mu$ the level density is finite.
On the contrary, if the two nucleons are moving with a finite center of mass momentum, also a hole develops at the Fermi level similar to what we have seen for the three-particle case. The width of this hole increases with increasing center of mass momentum until the gap disappears. This signals the critical  center of mass momentum. The finiteness of the level density at the Fermi level for two particles at rest is unique for the case of many-body level densities. This being the reason why pairing is such a unique phenomenon.

In conclusion, we have seen in this section that the density dependence of
alpha condensation in infinite matter is somewhat lower but still in line with the mean
field studies in finite nuclei presented above. This means in particular that
the action of the Pauli principle on the existence of alpha clusters is
similar in infinite matter and finite nuclei. 
As mentioned above the density of the Hoyle state is $\rho_0/3 - \rho_0/4$,
very close to the mean field values in this study. In~\cite{zho19} for the
hypothetical four-alpha condensate state in $^{16}$O at 15.1 MeV the
calculation yields quite a bit a lower density close to $\rho_0/6$. However,
the four-alpha calculation is more involved and the density of the
condensate may not be equivalent to the critical
density which can be higher. Also the four-alpha calculation may be
more sensitive to small perturbations like the increased (with
respect to the $^{12}$C case) Coulomb repulsion.


\section{Conclusions}

In conclusion, we have studied in nuclear systems the transition from a
Fermi gas to alpha-clustering as a function of density at zero
temperature. A first study based on dimensionless ratios of characteristic
length- and energy-scales of the nuclear many-body problem lead to a 
Mott-like transition from a delocalized phase to an alpha-clustered one
at the critical density $\sim\rho_0/5$, under which the Pauli principle
does not prevent anymore the formation of 4-nucleon bound states.
Very satisfactorily, these results are consistent with calculations 
in infinite matter, where the
phase transition from the Fermi-gas to alpha-particle condensation
happens at the critical density $\rho_\text{Mott} =
0.03 \text{fm}^3\sim \rho_0/5$~\cite{Sogo2,rop14}.
On the other hand, we also made constrained
HFB calculations, both with RMF and Gogny EDFs, where the radius of $^{16}$O is
increased under the constraint that on average the system stays spherical,
that is that the mean-value of the quadrupole operator remains zero. Also
in this way the system shows a critical radius, i.e., low density where
the homogeneously inflated $^{16}$O nucleus abruptly goes over into a
tetrahedral configuration of four alphas. This
happens consistently with the relativistic and non-relativistic approaches
at practically the same critical density  $\sim \rho_0/3$, slightly higher than
in the infinite
matter calculation as well as the analysis of dimensionless parameters of the nuclear many-body problem. 
This shows that the Pauli-principle which triggers
this QPT-transition acts rather similarly independently of whether the system
goes over into a condensate or a lattice configuration as this happens
with the constrained mean field calculation for $^{16}$O. We further investigated
the transition to four alphas in $^{16}$O. We expressed the non-dynamical 
correlations grasped through the SSB of the $O(3)$ rotation group down to the point
group $Td$ in terms of ph excitations on top of a symmetry-preserving SD, and discussed 
the crucial role played by orbitals beyond the p- and sd-shells to localize nucleons 
into alphas at the corners of a tetrahedron. All in all, $^{16}$O provides a rather spectacular example of a quantum phase transition in nuclear physics.

\begin{acknowledgments}
One of us (P.S) wishes to thank G. Roepke for useful discussions.
This publication is based on work supported in part by the framework 
of the Espace de Structure et de r\'eactions Nucl\'eaires Th\'eorique (ESNT) at CEA.
\end{acknowledgments}


\end{document}